\documentclass[a4paper,11pt]{article}
\pdfoutput=1 

\usepackage{jinstpub} 
\usepackage{siunitx}
\usepackage{color}

\title{Design and Sensitivity of the Radio Neutrino Observatory in Greenland (RNO-G)}
\author[1]{J.~A.~Aguilar}
\author[2]{P.~Allison}
\author[2]{J.~J.~Beatty}
\author[3]{H.~Bernhoff}
\author[4,5]{D.~Besson}
\author[6]{N.~Bingefors}
\author[6]{O.~Botner}
\author[7]{S.~Buitink}
\author[8]{K.~Carter}
\author[9]{B.~A.~Clark}
\author[2]{A.~Connolly}
\author[1]{P.~Dasgupta}
\author[10]{S.~de Kockere}
\author[10]{K.~D.~de Vries}
\author[11]{C.~Deaconu}
\author[12]{M.~A.~DuVernois}
\author[13]{N.~Feigl}
\author[13,14]{D.~Garc\'{\i}a-Fern\'{a}ndez}
\author[6]{C.~Glaser}
\author[6]{A.~Hallgren}
\author[14]{S.~Hallmann}
\author[15]{J.~C.~Hanson}
\author[17]{B.~Hendricks}
\author[12]{B.~Hokanson-Fasig}
\author[4]{C.~Hornhuber}
\author[11]{K.~Hughes} 
\author[12]{A.~Karle}
\author[12]{J.~L.~Kelley}
\author[16]{S.~R.~Klein}
\author[17]{R.~Krebs}
\author[13]{R.~Lahmann}
\author[4]{M.~Magnuson}
\author[12]{T.~Meures}
\author[13,14]{Z.~S.~Meyers}
\author[14,13]{A.~Nelles}
\author[4]{A.~Novikov}
\author[11]{E.~Oberla} 
\author[18]{B.~Oeyen}
\author[7]{H.~Pandya}
\author[13,14]{I.~Plaisier}
\author[19,14]{L.~Pyras}
\author[18]{D.~Ryckbosch}
\author[10]{O.~Scholten}
\author[20]{D.~Seckel}
\author[11]{D.~Smith} 
\author[11]{D.~Southall} 
\author[2]{J.~Torres}
\author[1]{S.~Toscano}
\author[10,7]{D.~J.~Van Den Broeck}
\author[10]{N.~van Eijndhoven}
\author[11]{A.~G.~Vieregg}
\author[13,14]{C.~Welling}
\author[17, 8]{S.~Wissel}
\author[4]{R.~Young}
\author[13]{A.~Zink}

\affiliation[1]{Universit\'e Libre de Bruxelles, Science Faculty CP230, B-1050 Brussels, Belgium}
\affiliation[2]{Dept.~of Physics, Center for Cosmology and AstroParticle Physics, Ohio State University, Columbus, OH 43210, USA}
\affiliation[3]{Uppsala University, Dept.~of Engineering Sciences, Division of Electricity, Uppsala, SE-752 37, Sweden}
\affiliation[4]{University of Kansas, Dept. of Physics and Astronomy, Lawrence, KS 66045, USA}
\affiliation[5]{National Nuclear Research University MEPhI, Kashirskoe Shosse 31, 115409, Moscow, Russia}
\affiliation[6]{Uppsala University, Dept.~of Physics and Astronomy, Uppsala, SE-752 37, Sweden}
\affiliation[7]{Vrije Universiteit Brussel, Astrophysical Institute, Pleinlaan 2, 1050 Brussels, Belgium}
\affiliation[8]{Physics Dept. California Polytechnic State University, San Luis Obispo CA 93407, USA}
\affiliation[9]{Dept.~of Physics and Astronomy, Michigan State University, East Lansing MI 48824, USA}
\affiliation[10]{Vrije Universiteit Brussel, Dienst ELEM, B-1050 Brussels, Belgium}
\affiliation[11]{Dept.~of Physics, Enrico Fermi Inst., Kavli Inst.~for Cosmological Physics, University of Chicago, Chicago, IL 60637, USA} 
\affiliation[12]{Wisconsin IceCube Particle Astrophysics Center (WIPAC) and Dept.~of Physics, University of Wisconsin-Madison, Madison, WI 53703,  USA}
\affiliation[13]{Erlangen Center for Astroparticle Physics (ECAP), Friedrich-Alexander-University Erlangen-Nuremberg, 91058 Erlangen, Germany}
\affiliation[14]{DESY, Platanenallee 6, 15738 Zeuthen, Germany}
\affiliation[15]{Whittier College, Whittier, CA 90602, USA} 
\affiliation[16]{Lawrence Berkeley National Laboratory, Berkeley, CA 94720, USA}
\affiliation[17]{Dept.~of Physics, Dept.~of Astronomy \& Astrophysics, Penn State University, University Park, PA 16801, USA}
\affiliation[18]{Ghent University, Dept. of Physics and Astronomy, B-9000 Gent, Belgium}
\affiliation[19]{Humboldt-Universit\"at zu Berlin, Unter den Linden 6, 10117 Berlin, Germany}
\affiliation[20]{Dept.~of Physics and Astronomy, University of Delaware, Newark, DE 19716, USA}

\emailAdd{anna.nelles@desy.de, authors@rno-g.org}

\abstract{This article presents the design of the Radio Neutrino Observatory Greenland (RNO-G) and discusses its scientific prospects. Using an array of radio sensors, RNO-G seeks to measure neutrinos above \SI{10}{PeV} by exploiting the Askaryan effect in neutrino-induced cascades in ice. We discuss the experimental considerations that drive the design of RNO-G, present first measurements of the hardware that is to be deployed and discuss the projected sensitivity of the instrument. RNO-G will be the first production-scale radio detector for in-ice neutrino signals.}

\begin{document}

\maketitle
\flushbottom

\section{Introduction}

This paper describes the Radio Neutrino Observatory Greenland (RNO-G) as it will be constructed at Summit Station in Greenland starting in 2021. RNO-G science targets astrophysical neutrinos of several PeV in energy up to the EeV range.

In this paper, we first motivate the science case for RNO-G, elaborate on experimental design considerations and then outline the instrument design. Awaiting in-field performance data, this article does not serve as a technical document, but describes the concept, the current hardware developments and boundary conditions behind the RNO-G approach. We conclude with a description of initial estimates of the design sensitivity of the instrument, as well as the expected resolution for such quantities as neutrino arrival direction and energy.

\subsection{Scope of RNO-G}

RNO-G will be constructed over three installation seasons. RNO-G will reach unprecedented yearly sensitivity to neutrino signals above 10 PeV, and will demonstrate a large-scale implementation (35 stations) of the in-ice radio neutrino detection technique. Even further scaling up of the in-ice radio technique, beyond the scale of RNO-G, is being developed as part of IceCube-Gen2 \cite{Aartsen:2020fgd}.

Considering both logistical constraints and also science opportunities (detailed below), RNO-G will be constructed at Summit Station in Greenland.  The RNO-G collaboration consists of members of all previous radio in-ice neutrino experiments from both Europe and the United States.  

\subsection{Relation to previous and current radio experiments}

Due to the extremely low neutrino flux at energies above 10 PeV, no neutrino has yet been detected using the radio technique. However, several experiments have shown the feasibility of this detection method and its potential.
RNO-G builds heavily on the experience of previous radio neutrinos detectors, like the pioneering RICE \cite{rice03,rice06}, the ARA \cite{ara_2station,ara_performance,ara2019} and ARIANNA \cite{arianna15a,arianna15b} experiments, as well as the balloon-borne ANITA \cite{anita06,anita10} experiment. These efforts tested different aspects of the radio technique and helped illuminate technologically important aspects of operating in remote locations in harsh polar conditions. 

The first experience with in-ice radio detectors was gained with the Radio Ice Cherenkov Experiment (RICE) \cite{rice03} at the South Pole. After a number of prototypes and initial measurements of the ice characteristics, the main experiment operated from 1999 until 2010. RICE provided the first neutrino limits \cite{rice06} from radio detectors and valuable experience in operating radio detectors at depths of down to \SI{200}{m}.

The Askaryan Radio Array (ARA)~\cite{ara_performance} has operated at South Pole since 2010~\cite{ara_iceray} and is a direct successor to RICE. While the RICE antennas were co-located with the AMANDA and IceCube experiments at South Pole, all five ARA stations operate in dedicated dry holes of depths \SIrange{50}{200}{m}. While different hardware has been deployed in different ARA stations, the station layout is mostly uniform. Every station consists of four receiver strings down to \SI{200}{m}. Each string is equipped with two vertically-polarized birdcage dipole antennas (VPol) and two ferrite-loaded slot antennas (Hpol) to reconstruct the radio signals. In addition, one or two calibration strings as well as surface antennas (on the earlier stations) are deployed. As the narrow cylindrical borehole geometry limits the intrinsic antenna gain, ARA pioneered the phased-array technique for radio detection of neutrinos at the most recently completed station \cite{oberla}.  

To date, the ARA collaboration has published constraints on the diffuse ultra-high energy (UHE) neutrino flux~\cite{ara2019}, neutrinos from gamma-ray bursts (GRBs) \cite{ara_grb}, and radio emission from solar flares~\cite{brian-amypaper}. The performance of the instrument has been verified using transmitters lowered into the SPICE borehole \cite{casey_2014}, which also allowed for the measurement of glaciological properties of the ice -- some of which can be used for improved neutrino event reconstruction \cite{glaciology-draft,Allison:2019rgg}. 

The Antarctic Ross Ice-Shelf ANtenna Neutrino Array (ARIANNA) began construction at the Ross Ice-Shelf in 2010, with a first hexagonal radio array being completed in 2015 \cite{arianna15a,arianna15b}. The ARIANNA concept is based around surface stations, i.e. the antennas are deployed just underneath the snow-surface. High-gain log-periodic dipole antennas (LPDAs) are deployed in shallow slots in the snow, where they are not restricted by the borehole geometry and exhibit broadband characteristics and dedicated polarization sensitivity, particularly to horizontally polarized signals. By placing the antennas at Moore's Bay on the Ross Ice-Shelf, the neutrino-detection strategy utilizes the reflective surface at the bottom of the ice at the water interface, which reflects downward going neutrino signals back to the stations. Without external infrastructure, ARIANNA pioneered autonomous low-power stations, based on renewable energy sources, operated via wireless communications. Most recently wind turbines were added to the solar power-provision system \cite{WindICRC}. 

ARIANNA has successfully detected the radio signal of air showers as calibration and verification signals \cite{Barwick17} and published limits on the UHE neutrino flux \cite{Anker:2019rzo}. The collaboration also published the effectiveness of recording signals reflected from the surface by monitoring snow accumulation \cite{Anker:2019zcx}. Two ARIANNA stations have also been deployed at South Pole to test the robustness of the hardware under environmental circumstances differing from the Ross Ice Shelf. The same calibration source as used for ARA from the SPICE borehole was then also used to verify the reconstruction capabilities of the ARIANNA experiment with respect to arrival direction and polarization \cite{Anker:2020bjs}. 

The Antarctic Impulsive Transient Antenna (ANITA) experiment has flown four separate missions over Antarctica. ANITA is a balloon-borne radio receiver array that scans the surface from afar for upcoming neutrino signals generated below the ice surface. Several components of the ANITA hardware have been incorporated into the ARA and ARIANNA designs \cite{anita06,anita10,anita3}. While equipped with much different power and lifetime requirements, the technological challenges remain similar. A data acquisition system with high timing accuracy and thorough calibration is needed to reliably reconstruct neutrino or cosmic-ray signals. ANITA was the first radio-neutrino experiment to report the detection of air shower signals \cite{Hoover:2010qt}, which helped to verify the simulation chain and the understanding of the energy calibration \cite{anita_CRs}. The ANITA collaboration observed several events which, if neutrinos, would seem to be in tension with Standard-Model cross-sections \cite{anita1me,anita3me, Romero-Wolf:2018zxt}. Those events may also stem from unexplained systematics or ice effects \cite{deVries:2019gzs,shoemaker2020}.

Operation of existing ARA stations continue in close cooperation with IceCube. In addition, proposals for an ANITA-successor ballooning effort are being discussed, as well as an extension of the ARIANNA array at Moore's Bay.  

In addition to building on experiences with dedicated radio neutrino experiments, RNO-G also profits from knowledge gained at accelerator experiments about the nature of the in-medium emission from particle showers \cite{Saltzberg:2000bk,Gorham:2004ny,Gorham:2006fy,t510}, as well as those from mid-scale air shower arrays measuring the radio emission of cosmic ray induced showers e.g.\ \cite{Falcke:2005tc,Ardouin:2009zp,Aab:2016eeq,Schellart:2013bba,Bezyazeekov:2015ica}. First efforts at exploring the feasibility of a detector in Greenland have been conducted previously by members of the collaboration \cite{Wissel:2015eqm,avva1,Deaconu:2018bkf} and have encouraged the development of RNO-G.

\section{Science case and design requirements}
Neutrinos are ideal messengers to identify the UHE sources in the universe. Unlike cosmic rays, which are deflected by magnetic fields and interact with intervening matter and radiation, neutrinos point back to their sources and can reach Earth from the most distant corners of the universe. Furthermore, due to their low interaction cross section, neutrinos are unique messengers to convey information about the inner engine of cosmic accelerator sites. Unlike $\gamma$-rays, which can also be created by inverse Compton scattering, the observation of high-energy neutrinos from astronomical objects provides incontrovertible evidence for hadronic cosmic-ray acceleration.
Identifying the sources of cosmic rays and the acceleration mechanisms requires a comprehensive multi-messenger observation program comprising cosmic rays, $\gamma$-rays, and neutrinos across many decades of energy.

\begin{figure}
\centering
\includegraphics[width=\textwidth]{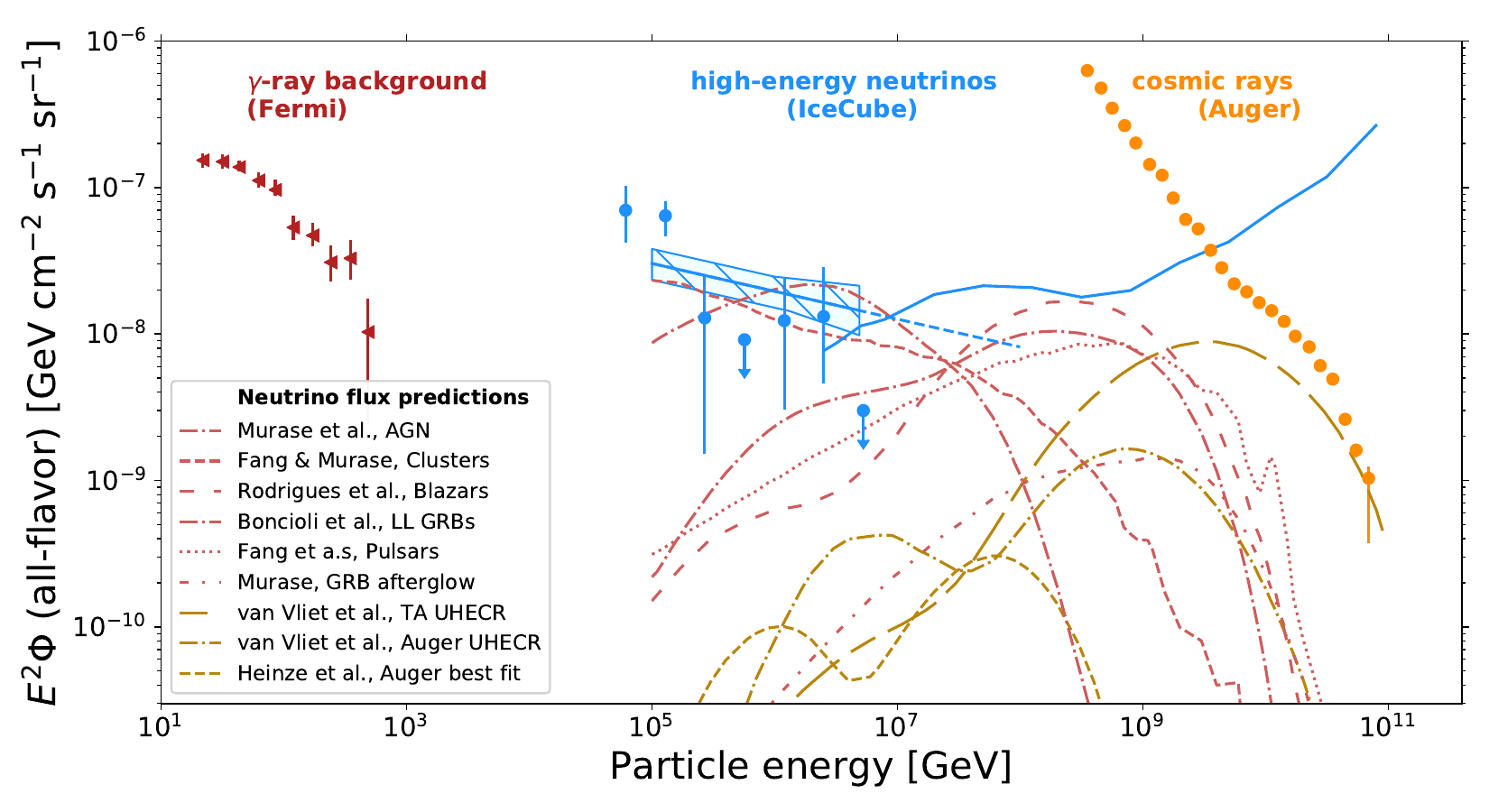}
\caption{A multi-messenger view of the high-energy universe, inspired by \cite{ahlers_mm_figure_source}, showing the science reach for radio detection of neutrinos. Shown are models predicting neutrinos from sources (in red lines) \cite{Fang:2017zjf,FangPulsar,Boncioli:2018lrv,Murase2014AGNJets,rodrigues2020blazar,Murase:2007yt} and those from the interaction of the ultra-high energy cosmic rays with various photon backgrounds (in dark yellow lines). Overlaid are \cite{Heinze:2019jou,vanVliet:2019nse} the $\gamma$-ray measurements from Fermi \cite{Ackermann:2014usa}, the IceCube neutrino measurements and the fit to the muon neutrino spectrum \cite{Haack:2017dxi,Kopper:2017zzm,Aartsen:2018vtx}, as well as the spectrum of ultra-high energy cosmic rays as reported by the Pierre Auger Observatory \cite{Aab:2015bza}. 
}
\label{fig:sensitivity}
\end{figure}

In the last years, neutrinos have delivered on their promise to provide a key piece of this astronomical puzzle with the discovery of a diffuse flux of astrophysical neutrinos~\cite{IceCube2013a,IceCube2014a,IceCube2015a,IceCube2016a}. 
IceCube has measured the neutrino energy spectrum to above \SI{1}{PeV} -- the highest-energy neutrinos ever observed.  Beyond the PeV scale, the limited size of IceCube prohibits observation of the steeply falling neutrino flux.
Fig.~\ref{fig:sensitivity} compares the neutrino flux measured by IceCube with the diffuse flux of $\gamma$-rays measured by Fermi~\cite{Ackermann:2014usa} and the cosmic-ray spectrum  measured by Auger~\cite{Aab:2015bza}. The three spectra display tantalizingly similar energy densities, suggesting a common origin. In such a scenario, cosmic-ray collisions produce pions, where gamma-rays then stem from decays of neutral pions and neutrinos from those of charged pions. The figure also shows the gap in observations of UHE neutrinos beyond the energies reachable by IceCube. 

Multi-messenger observations are even more intriguing in light of the announcement in July 2018 of the first coincident observation of a neutrino from the direction of a source (the blazar TXS 0506+056) that was flaring simultaneously in $\gamma$-rays ~\cite{IceCube2018a, IceCube2018b}.  
This was also the first multi-messenger observation triggered by a high-energy neutrino, demonstrating the capability to send real time alerts and establishing the field as a vital pillar of multi-messenger astronomy. To fully understand the neutrino sky, however, a larger detector must be built and observations extended to the PeV--EeV energy range.

The radio detection technique naturally targets neutrino energies beyond the reach of IceCube. Due to the kilometer scale attenuation length of radio waves in ice, very sparse radio detectors cover large volumes of material, providing huge effective volumes at \SI{10}{PeV} to \SI{100}{EeV}. In this energy range, several transient and diffuse sources of neutrinos are expected and an experimental measurement would strongly impact identification of the sources of ultra-high energy cosmic rays. 

The general science case of neutrino astronomy has been reviewed in the context of the 2020 US decadal survey \cite{Astro2020_nuastro,Astro2020_nufundamental}. This section will thus focus specifically on the science program that can be conducted by radio detectors for high-energy neutrinos. 

\subsection{Diffuse neutrino flux}

The radio detection of neutrinos targets the energy range from \SI{10}{PeV} to beyond \SI{100}{EeV}. In this range, diffuse neutrino fluxes both directly from sources (\textit{astrophysical neutrinos}), as well as from the interaction of ultra-high energy cosmic rays (UHECRs) with photon backgrounds (\textit{cosmogenic neutrinos}) are predicted. Detecting either will enable studies of high-energy neutrino production mechanisms locally, at the still unknown sources.

Fig.~\ref{fig:sensitivity} shows different models for astrophysical (red) and cosmogenic (yellow) neutrinos that fall in the energy range of radio detectors.
Cosmogenic neutrinos result from interactions of UHECRs with photon fields like the extra-galactic background light, the infra-red background, or the cosmic microwave background \cite{BerezinskyZ}. The flux and spectrum of these neutrinos are grounded in the UHECR mass composition, but are subject to model assumptions about the cosmological luminosity and chemical evolution of the sources, which can differ outside of the local universe probed by UHECRs~\cite{Ahlers2012}. For the cosmogenic neutrino predictions shown in Fig.~\ref{fig:sensitivity}, we compare predictions based on compositions measured by the Telescope Array (TA)~\cite{TA_composition, TA_composition_1} and the Pierre Auger Observatory (Auger) \cite{Heinze:2019jou,auger16}. These are in fact only examples of the full range of possible models admitted by current constraints \cite{vanVliet:2019nse}. 

While the cosmogenic fluxes predicted assuming the Auger and TA compositions vary significantly, composition measurements from the two experiments are
compatible within systematic uncertainties \cite{Deligny:2020gzq}. With a measurement of UHE neutrinos, radio detectors can resolve the question of a pure-proton composition, which is disfavored by Auger, but still allowed by TA data. More generally, measuring UHE neutrinos will constrain a combination of proton fraction, source evolution and highest-energy cutoffs of UHECRs well beyond local sources.

We consider `astrophysical' neutrinos as those created directly in (or very close to) the sources of UHECRs. These neutrinos tend to have lower energies than cosmogenic neutrinos, but also reach the energy range of radio detectors. They will definitely trace their sources, allowing for stacking analyses to reveal them. These neutrinos are not necessarily time-coincident with explosive events (see Sect.~\ref{sec:transients}), but contribute to a constant diffuse flux. Potential candidates range from Active Galactic Nuclei (AGN) \cite{Murase2014AGNJets} to various types of gamma-ray bursts (GRBs) \cite{Boncioli:2018lrv,Murase:2007yt}, pulsars \cite{FangPulsar}, galaxy clusters \cite{Fang:2017zjf}, Flat Spectrum Radio Quasars (FSRQs) \cite{righi2020eev}, and blazars \cite{rodrigues2020blazar}. 

The diversity of models of astrophysical neutrinos is already large and promising, but we expect more models to become available as detectors with the necessary sensitivities are commissioned.

It remains to be explored whether astrophysical neutrinos are the source of the diffuse flux as measured by IceCube or whether the observed flux is the low energy tail of the cosmogenic neutrinos. So far, despite the multi-messenger successes, studies demonstrate that neutrinos from blazars cannot comprise the bulk of the diffuse neutrino spectrum at energies accessible by IceCube \cite{Murase:2016gly,Ando:2017xcb,Aartsen:2016lir,Neronov:2016ksj,Huber:2017wxt,Hooper:2018wyk}. A radio detector will be able to measure the continuation of the IceCube flux to higher energies and thereby provide additional information on the spectral shape of the flux, which may be useful to disentangle the source contributions. 

A successful search for the diffuse neutrino flux at energies beyond \SI{10}{PeV} requires, above all, an adequate flux sensitivity to ensure a first observation. To subsequently discriminate putative production mechanisms, a detector must provide an adequate energy estimate for every neutrino and an angular reconstruction that allows for the correlation of arrival directions with known sources. 

\subsection{Sky coverage}
\label{ref:sky_coverage}

\begin{figure} 
\includegraphics[width=\textwidth]{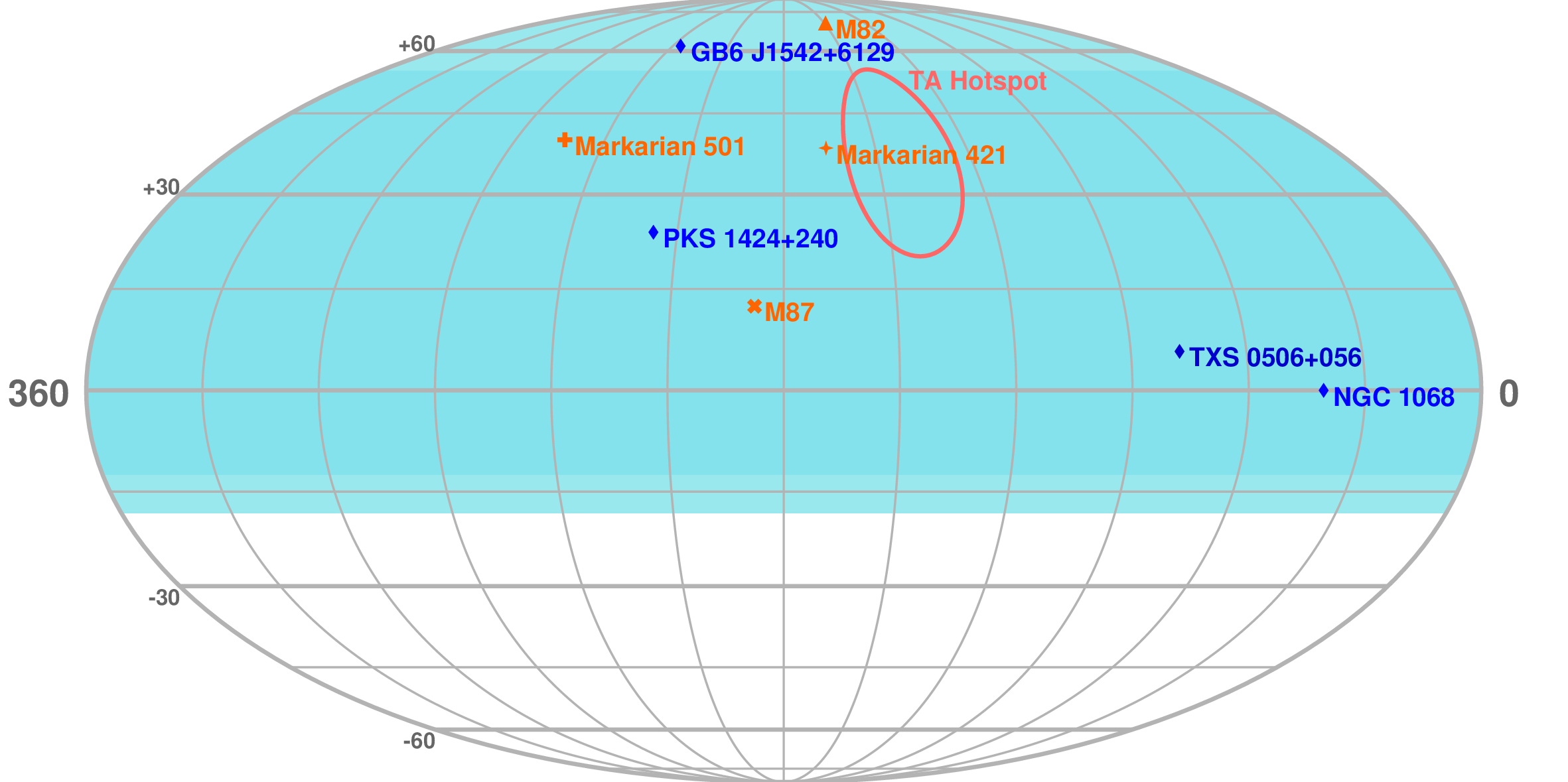}
\caption{The field of view, in equatorial coordinates, of an in-ice radio detector for neutrinos in Greenland. The colored background represents the diurnally-averaged total field of view of the detector. Also shown are targets with interesting multi-messenger implications. The blue sources are those seen by IceCube as the most significant sources in a point-source search \cite{Aartsen:2019fau}. In orange, we show other interesting candidates, with strong $\gamma$-ray emission and/or radio emission. Furthermore, we indicate what is known as the \textit{TA hotspot} as indicated by the anisotropy measurement in cosmic ray measured with the Telescope Array \cite{Abbasi:2014lda}. 
} 
\label{fig:skymap}
\end{figure}

Fig.~\ref{fig:skymap} demonstrates the field of view of a radio neutrino telescope sited in Greenland. When targeting point-like sources, either steady or transient (see Sec.~\ref{sec:transients}), the field of view of the detector becomes relevant. The Earth is opaque to neutrinos at PeV to EeV energies, such that UHE neutrino observatories are most sensitive to down-going or Earth-skimming neutrinos. As will be discussed in more detail Sec.~\ref{sec:emission}, a radio neutrino detector in glacial ice on bedrock will be most sensitive to an annulus above the horizon. 

Combining the opacity of the Earth to neutrinos above PeV energies with the inherent radio detector sensitivity means that, for example, a follow-up of TeV-scale IceCube events at higher energies requires a Northern detector such as RNO-G. A \textit{single event} observed by a radio detector in the Northern hemisphere will define the flux in a new energy regime, and even a non-detection will constrain the allowed flux through \textit{multi-wavelength} neutrino observations.

The continuous sky coverage and large field-of-view will enable studies of point sources of high-energy neutrinos. The hotspot of UHECRs observed by TA~\cite{Abbasi:2014lda} (red ellipse in Fig.~\ref{fig:skymap}) lies in the Northern Hemisphere. While the cosmogenic neutrino flux is expected to be diffuse, studies attributing the TA hotspot to a single source of cosmic rays like M82 predict point sources of EeV neutrinos~\cite{TAHotSpot_M82}.  There are additionally four intriguing point sources nearing the threshold for a high-confidence long-term detection in IceCube (shown as navy blue diamonds in Fig.~\ref{fig:skymap}), all of which lie in the Northern Hemisphere due to the sensitivity of IceCube. These include not only TXS 0506+056, but also NGC 1068, an AGN which lies near the strongest hotspot in IceCube's all-sky scan~\cite{Aartsen:2019fau}.

\subsection{Transient sources}
\label{sec:transients}

Detecting neutrino emission in temporal and spatial coincidence with an explosive event has shaped and will continue to shape multi-messenger astronomy~\cite{IceCube2018a, IceCube2018b}. By uniquely identifying sources, neutrinos will help to characterize and discover the most energetic non-thermal sources on the sky. Many models of astrophysical transient phenomena predict neutrinos in the detectable energy range of radio neutrino detectors. 

\begin{figure}
    \centering
    \includegraphics[width=0.6\textwidth]{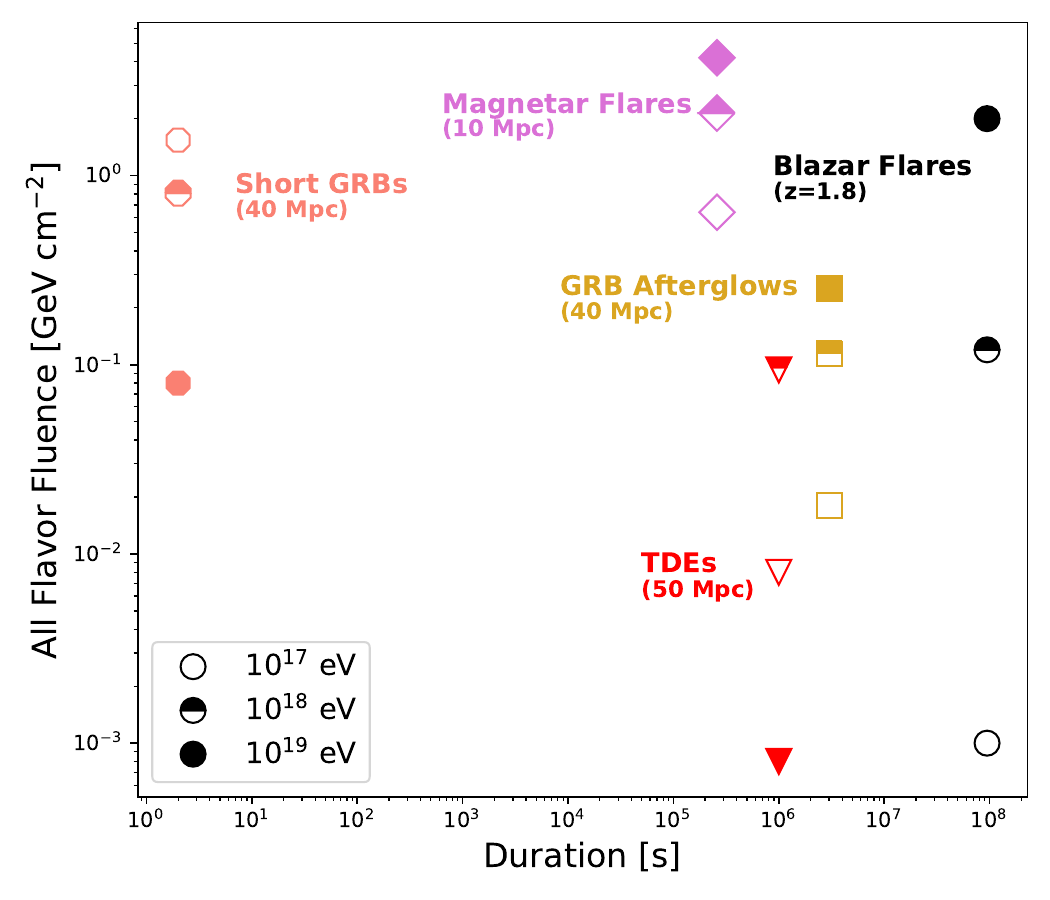}
    \caption{Expected neutrino flux at Earth integrated over transient duration from different classes of astrophysical objects placed at probable and previously studied distances. These include short GRBs (possibly associated with a Neutron Star-Neutron Star merger) \cite{Kimura:2017kan} and GRB afterglows \cite{Murase:2007yt} at a distance of 40 Mpc, the high state SED of Tidal Disruption Events (TDEs) at 50 Mpc \cite{Guepin:2017abw}, and stable fast-spinning young magnetars \cite{Fang:2017tla} at 10 Mpc roughly 4 days post merger. Blazar flares \cite{rodrigues2020multi} are shown based on a proton synchotron model applied to multi-wavelength observations of PKS 1502+106, which has a redshift of z=1.8.  The expected transient fluence for each object is plotted against duration, or flare time, for three neutrino energies: $10^{17}$ eV (hollow shapes), $10^{18}$ eV (half shapes), and $10^{19}$ eV (filled shapes).   
    }
    \label{fig:duration}
\end{figure}

The overlap in sky coverage with IceCube, where IceCube has its best efficiency for directional reconstruction of astrophysical neutrinos, will enable studies of several interesting flaring, transient sources over a broad energy band. Should the first tentative extra-galactic neutrino source, the blazar TXS 0506+056, flare~\cite{IceCube2018a, IceCube2018b} again, observations made by IceCube and RNO-G may be able to define the neutrino spectrum. Similarly, the first blazars known to flare with TeV $\gamma$-rays emission, Markarian 501~\cite{Mrk501ApJ1996} and Markarian 421~\cite{Mrk421Nature1992}, also lie in the Northern sky. Models of transient bursts of neutrinos due to tidally disrupted stars~\cite{Wang:2015mmh,Dai:2016gtz,Senno:2016bso,Lunardini:2016xwi,Zhang:2017hom,Biehl:2017hnb,Guepin:2017abw} and binary neutron star mergers~\cite{Metzger, GWBNS} also predict neutrinos in the PeV to EeV energy scale. The latter are targets for multi-messenger observations of gravitational waves and neutrinos. Fig.~\ref{fig:duration} shows a fraction of the parameter space over which neutrinos are expected as transient phenomena from various source classes. In the figure, model-dependent fluence is compared to duration for varying neutrino energies around EeV.   Furthermore, different populations of blazars, including low-luminosity BL Lacs, high-luminosity BL Lacs and FSRQs \cite{rodrigues2020blazar}, the most powerful blazars in the $\gamma$-ray band \cite{righi2020eev}, could provide intriguing candidates for multi-wavelength follow up. The energy threshold of RNO-G will allow sensitive searches for GRBs \cite{Waxman:1997ti,Rachen:1998fd,Dermer:2003zv,Guetta:2003wi,Razzaque:2003uw,Murase:2005hy,Murase:2008sp,Wang:2008zm,Baerwald:2010fk,Ahlers:2011jj,Murase:2011cx,Li:2011ah,Hummer:2011ms,He:2012tq,Zhang:2012qy,Liu:2012pf,Gao:2013fra,Petropoulou:2014awa,Petropoulou:2014lja,Bustamante:2014oka,Wang:2007xj,Murase:2008mr,Calvez:2010uh,Globus:2014fka,Biehl:2017zlw,Paczynski:1994uv,Bartos:2013hf,Murase:2013hh,Murase:2006dr,Waxman:1999ai,Dermer:2000yd,Murase:2007yt,Razzaque:2013dsa,Murase:2006mm,Gupta:2006jm,Murase:2008mr,Senno:2015tsn,Zhang:2017moz,Boncioli:2018lrv,Zhang:2018agl} with lower neutrino luminosity than previously conducted with radio neutrino experiments~\cite{ara_grb, anitaGRB}. 

A successful radio detector for transient signals needs reliable absolute timing and good angular reconstruction. Ideally, the angular reconstruction is both sufficiently rapid and accurate to allow meaningful alerts to be quickly sent to the multi-messenger community. Absolute timing is critical to the multi-messenger mission.

\subsection{Fundamental physics}

High-energy cosmic neutrinos uniquely probe fundamental particles and interactions in an uncharted and otherwise unreachable energy and redshift regime, as summarized in \cite{Astro2020_nufundamental}. 

The energy regime of neutrino radio detectors encompasses a relatively unmapped parameter space, helping to answer questions about the fundamental neutrino properties such as the behavior of neutrino cross-sections \cite{Connolly:2011vc,CooperSarkar:2011pa,Bertone:2018dse,Romero:2009vu,Hooper:2002yq,Klein:2013xoa,Ellis:2016dgb,Klein:2019nbu} and flavor mixing at high energies \cite{Bustamante:2015waa,Learned:1994wg,Arguelles:2015dca, Shoemaker:2015qul,Gonzalez-Garcia:2016gpq, Rasmussen:2017ert, Ahlers:2018yom}, or even whether neutrinos are stable in general \cite{Chikashige:1980qk, Gelmini:1982rr, Tomas:2001dh,Beacom:2002vi, Baerwald:2012kc, Shoemaker:2015qul, Bustamante:2016ciw, Denton:2018aml}. There is the chance to contribute to broader phenomenology such as the nature of dark matter \cite{Feng:2010gw,Beacom:2006tt,Yuksel:2007ac,Murase:2012xs,Feldstein:2013kka, Esmaili:2013gha, Higaki:2014dwa, Rott:2014kfa, Dudas:2014bca, Ema:2013nda, Zavala:2014dla, Murase:2015gea, Anchordoqui:2015lqa, Boucenna:2015tra, Dev:2016qbd, Hiroshima:2017hmy, Chianese:2017nwe}, the quest for the fundamental symmetries of nature, \cite{AmelinoCamelia:1997gz, Hooper:2005jp, GonzalezGarcia:2005xw, Anchordoqui:2005gj, Bazo:2009en, Bustamante:2010nq, Kostelecky:2011gq, Diaz:2013wia, Stecker:2014oxa, Stecker:2014xja, Tomar:2015fha, Ellis:2018ogq, Laha:2018hsh} and/or potential hidden interactions with cosmic backgrounds \cite{Lykken:2007kp, Ioka:2014kca, Ng:2014pca, Blum:2014ewa, Shoemaker:2015qul, Altmannshofer:2016brv, Barenboim:2019tux}.

Overall, for a radio detector to provide experimental data for fundamental physics experiments, the highest priority is to detect neutrinos with adequate statistics. After this is given, the accuracy of statements regarding fundamental physics will strongly depend on the accuracy of the reconstruction. The obtainable energy resolution directly impacts spectral measurements and the accuracy of energy dependent quantities in fundamental physics. For studies relying on for example the amount of matter traversed, angular resolution has a direct impact. Potential flavor sensitivity of radio experiments would be interesting to answer yet another set of fundamental physics questions. 

\subsection{Radio emission from neutrino interactions in ice and consequences for site selection}
\label{sec:emission}

The radio emission following a neutrino interaction stems from the \emph{Askaryan} effect \cite{Askaryan:1962hbi}. Postulated more than 50 years ago, the effect has been demonstrated in accelerator experiments in several dieletrics including ice \cite{Saltzberg:2000bk,Gorham:2004ny,Gorham:2006fy,t510}, as well as identified as secondary emission mechanisms in air showers \cite{Aab:2014esa,Schellart:2014oaa}.

Askaryan emission is caused by showers developing in a (dense) medium. Thus, a radio signal follows the interaction of neutrino of all flavors, as long as a particle shower is generated, both for hadronic and electromagnetic showers \cite{ConnollyV}. It is also possible to detect showers induced by catastrophic energy-losses of secondaries such as muons or taus \cite{Garcia-Fernandez:2020dhb}. The emission is a coherent effect, originating in the charge imbalance resulting from medium electrons either Compton scattering into the advancing shower or annihilating with shower positrons. With respect to the surrounding medium a net-negative charge is present in the shower front.

\begin{figure}
    \centering
    \includegraphics[width=\textwidth]{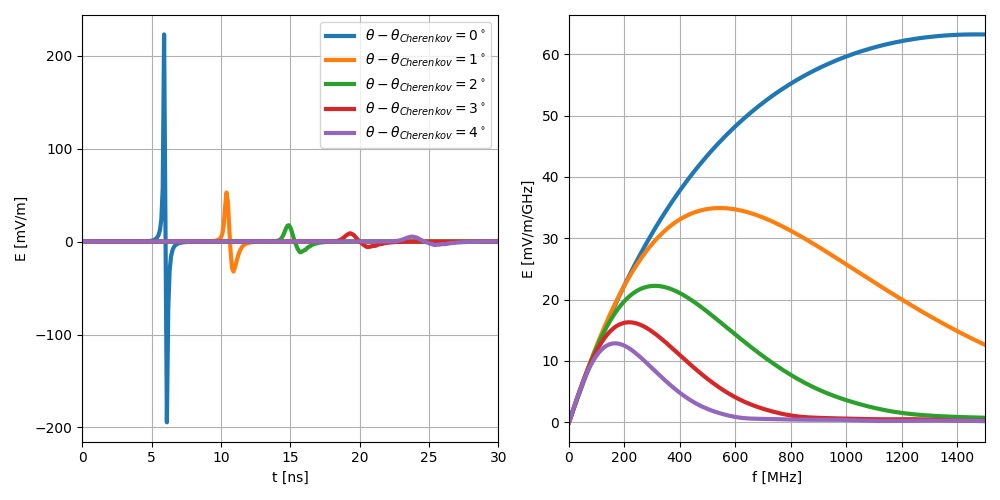}
    \caption{Electric-field waveforms (left) and spectra (right) of the radio signal emitted at different viewing angles relative to the Cherenkov angle, for a hadronic shower with energy deposition of 1 EeV. For enhanced readability, the waveforms have been offset in time. No propagation or detector effects have been included.
    }
    \label{fig:askaryan_pulses}
\end{figure}

The radio signal itself is a broad-band bipolar pulse with $\sim$ns-duration. Coherence is given over all frequencies (typically tens of MHz to tens of GHz) close to the Cherenkov angle, where the signal is strongest as the emission at all frequencies arrives in phase. Coherence is lost off the Cherenkov angle first at high frequencies, so that the Cherenkov ring is rather narrow at high frequencies and broader at low frequencies. A discussion of a variety of models for the radio emission of neutrinos can be found in \cite{NuRadioMC}, they range from simplified parameterized models in the frequency domain to more advanced semi-analytical time-domain models. In Fig.~\ref{fig:askaryan_pulses} we show typical pulses and their frequency spectra derived from \cite{ARZ}, for an illustration of the variety and the behavior. 

The energy threshold for a neutrino detection is significantly higher in radio than for optical instruments \cite{rice03}. Depending on the exact instrumental parameters, the pulse amplitude at a distance of \SI{100}{m} reaches the level of the typical thermal noise in low-noise radio receivers at approximately \SI{1}{PeV}. Although the energy per radio photon is significantly smaller than for optical photons, signal coherence compensates as the charge imbalance grows. As a coherent effect, the amplitude scales linearly with the number of excess electrons, which itself is linear in shower energy \cite{ZHS:1960, Saltzberg:2000bk}. However, it should be noted that the detected signal amplitude scales with $\frac{1}{r}$, with $r$ being the distance to the neutrino interaction vertex. 

At the same observer distance $r$, the detected signal amplitudes linearly as function of energy. This has been confirmed in air showers since the attenuation in air is negligible \cite{Aab:2016eeq}. The situation is different for instrumentation deployed in-ice. The kilometer-scale attenuation length in ice \cite{barrella,besson,ara_performance,avva1,Barwick05}, determines the range to an observable neutrino interaction, and, therefore, the detector effective volume. The attenuation length decreases with increasing temperature, which favors cold and thick ice for deployment. 

Naturally occurring ice follows a depth-dependent density profile with a gradient, from fresh snow to solid ice, resulting in a varying light velocity with depth, and therefore non-rectilinear ray trajectories.
In a medium with a refractive index gradient, radio signals are bent towards the denser medium, producing bent trajectories and a limited field of view for detectors in or close to the near-surface firn layer. These bent trajectories complicate the reconstruction, particularly when there are uncertainties in the ice properties. The simplest ansatz assumes a smooth ice density gradient. Calculations demonstrate that anisotropies in the firn (or below) may support unexpected horizontal propagation, as borne out by experimental data \cite{Deaconu:2018bkf,Barwick:2018rsp}. 
A radio detector should therefore preferably be built at a site with a small firn layer and otherwise smooth and homogeneous ice. 

Starting from PeV energies, the Earth is opaque to neutrinos, such that radio detectors will be sensitive to an annulus of neutrino directions above and slightly below the horizon. The deeper the detector, the more vertically incoming neutrino directions can be detected. For a detector at a few hundred meters depth, the sensitivity does not reach far beyond $30^{\circ}$ elevation, unless the reflective property of the bottom of a shelf-ice is used, as for the ARIANNA experiment \cite{Anker:2019mnx}. 

In summary, an in‐ice radio neutrino detector in glacial ice on bedrock will have the largest acceptance if installed in thick, smooth and cold ice. It will never be able to provide full sky coverage, but only be sensitive to a ring of elevations above and slightly below the local horizon. 

\begin{figure}
    \centering
    \includegraphics[width=0.9\textwidth]{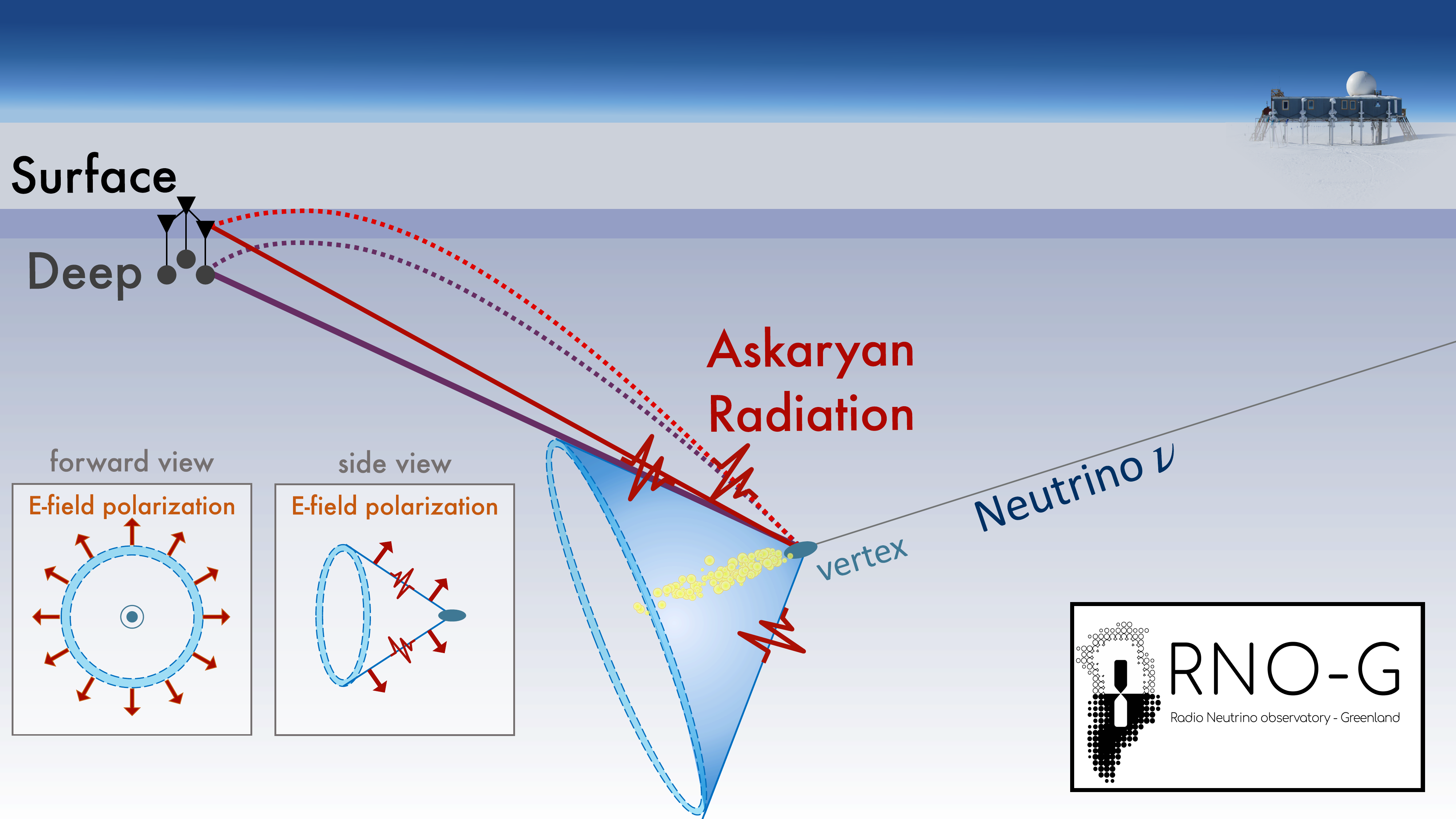}
    \caption{Schematic of the detection of the radio emission following a neutrino interaction (not to scale). The emission is strongest at the Cherenkov angle (blue cone) and can follow straight and bent trajectories to the receiving station depending on the profile of the index of refraction of the ice. The signal is usually detected at large distances and is strongly polarized as illustrated in the insets. 
    }
    \label{fig:schematic}
\end{figure}

Figure \ref{fig:schematic} provides an overview of the geometry for the detection radio signals with a detector buried in the ice. Every station monitors a large volume of ice, which means that by shear geometry a detection is most likely to show small signals as this corresponds to an interaction in the largest visible volume.

\subsection{Air showers as both a potential background and calibration signal}
The radio emission of air showers from the electron charge excess is similar to that for neutrino induced showers in ice. However, in air the geomagnetic emission \cite{KahnLerche1966,Allan:1972wd} dominates over the Askaryan effect. The geomagnetic emission stems from the charge separation induced by the Lorentz force in the Earth's magnetic field. 
The different signatures of the two contributions can be disentangled by their polarization. While still mostly linearly polarized, the main axis of the polarization from geomagnetic emission is aligned with the cross-product of shower axis and magnetic field \cite{Aab:2014esa,Schellart:2014oaa}. 

Due to their larger extent and the resulting consequences for coherence, air shower signals typically contain more low frequencies than those from showers in dense media \cite{Welling:2019scz}. Nevertheless, signals from air showers and denser in-ice showers are remarkably similar, which makes the much more abundant air shower signals a suitable calibration signal.
Since the cosmic ray energy spectrum is well-known (e.~g.~\cite{Deligny:2020gzq}) and the radio energy scale understood \cite{Aab:2016eeq,mulrey2020cosmicray}, measuring air showers will allow any detector to be calibrated {\it in-situ}, which includes checking the sensitivity simulations on an absolute scale. This will lend confidence to the signal identification and reconstruction \cite{Barwick17}. 

The remarkable similarity can of course also be a reason for concern. The in-air signal will be (partly) refracted into the ice, where it may be picked-up by antennas and incorrectly identified as neutrino induced signal. While the signal will clearly be down-going, so may be signals from neutrino interactions, due to the ray bending properties of the ice \cite{NuRadioMC}. It has also been argued that an incompletely developed air shower may cause transition radiation and other phenomena observable in deep detector stations \cite{deVries}. 
In addition, stochastic energy losses by high energy muons in an air shower penetrating the ice may mimic the interaction of a neutrino \cite{Garcia-Fernandez:2020dhb}. Without additional detectors, the muons themselves are invisible to radio detectors, while the energy losses are detectable. Depending on the exact detector configuration and trigger, these background events may limit the analysis efficiency, albeit dropping sharply in number with energy. 

\begin{figure}
\includegraphics[width=0.8\textwidth]{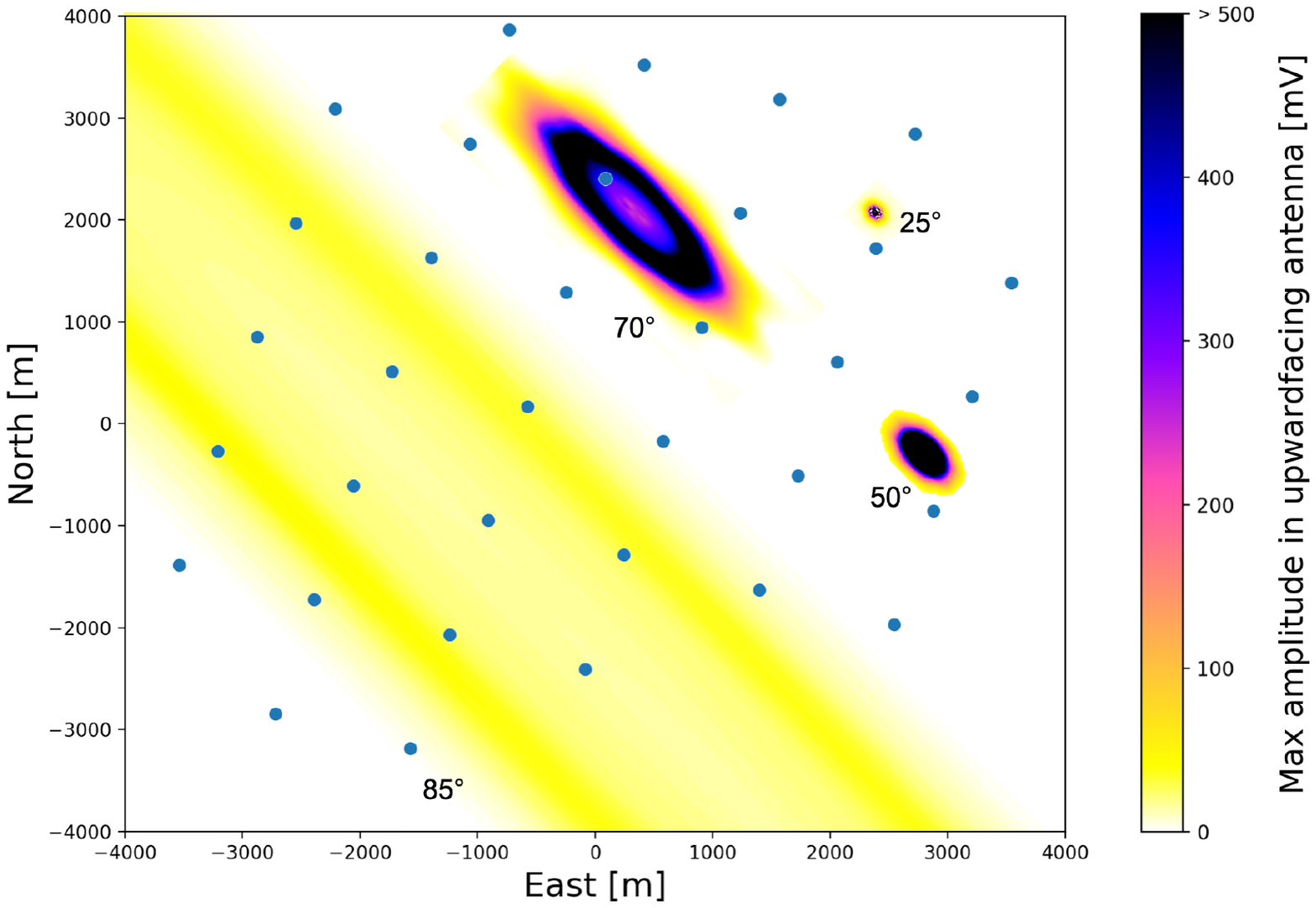}
    \centering
    \caption{Simulated radio air shower footprints at 4 different incoming zenith angles. Simulations were performed using CORSIKA and the RNO-G site (magnetic field and height above sea level) in Greenland, and include the response of upward facing logarithmic periodic dipole antennas, as planned for RNO-G (see Fig.~\ref{fig:layout}. The global maximum of the amplitudes in three antennas is shown. The air shower energy is \SI{3.2e18}{eV} for all showers, and the zenith angles are indicated in the figure. 
    }
    \label{fig:rno_airshower}
\end{figure}

Overall, this argues to equip all radio neutrino detectors with their own dedicated air shower array, for both calibration and veto purposes. Conveniently, due to the signal similarity, no additional technology is needed for such a detector, but does require additional surface antennas connected to the same data-acquisition system (DAQ). A dedicated air shower trigger, optimized to the lower frequency content of the air shower signals, would significantly enhance efficiency and detection rate. Due to the height of the interaction in the atmosphere and the fact that ${\rm n}_{air}\approx 1.0$ co-aligns the emission with the shower axis, the detectable footprint of the radio signal from air showers is centered on the shower axis, with lateral extent distributed ellipsoidally on the ground, as shown in Fig.~\ref{fig:rno_airshower}. The exact size is governed by the distance to shower maximum and the projection effect of the zenith angle \cite{Nelles:2014xaa}. The figure qualitatively illustrates that vetoing horizontal air showers will be relatively straightforward, while retaining high efficiency for vertical showers presents more of a challenge. The typical threshold for air shower detection is around \SI{10}{PeV}, which is again similar to the threshold of in-ice detection.


\section{Experimental design considerations}

\begin{figure} 
\includegraphics[width=0.48\textwidth]{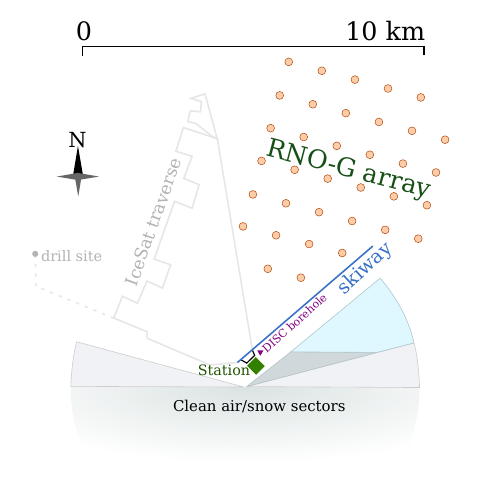}
\includegraphics[width=0.5\textwidth]{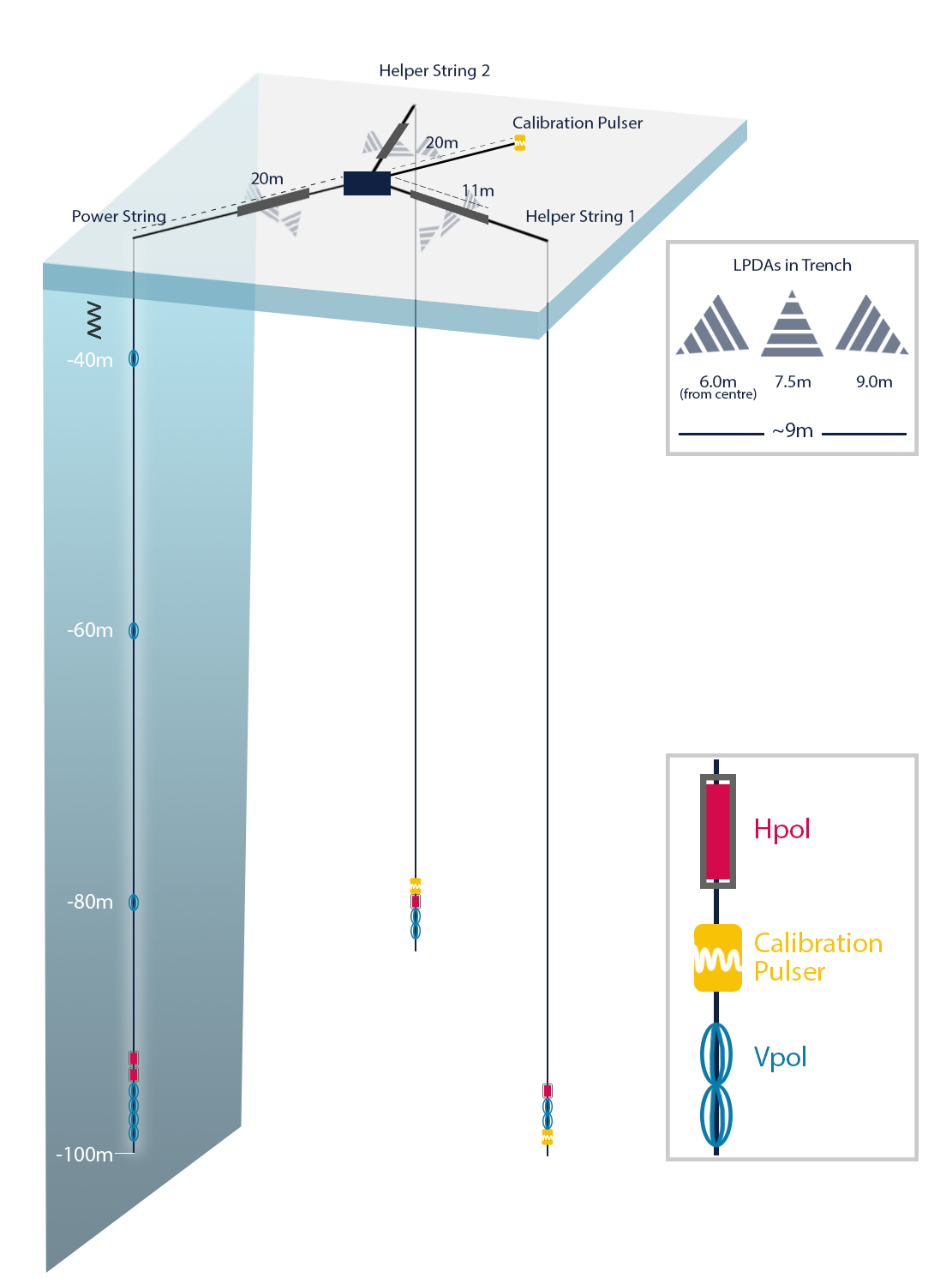}
\caption{Left: Map of the planned RNO-G array at Summit Station; grid spacing is approximately 1 km.  Right: A single RNO-G station consists of three strings of antennas (Hpol and Vpol) plus surface antennas (LPDAs), as well as three calibration pulsers located both deep in the ice and also at the surface. The string containing the phased array trigger is designated as the \emph{power string}, while the two additional strings are designated as \emph{support strings}. 
} 
\label{fig:layout} 
\end{figure} 

RNO-G is designed to demonstrate the scalability of the radio detection technology, while enabling the world's-best UHE neutrino sensitivity through low thresholds and also high efficiency. The system is designed to provide high fidelity identification of neutrino signals and reconstruction of neutrino properties. Building on these requirements, a station and array design as schematically depicted in Fig.~\ref{fig:layout} was developed.

The design of RNO-G combines the experience gained with all prior in-ice radio neutrino experiments, especially ARA \cite{ara_performance} and ARIANNA \cite{Barwick:2015ica}, and also builds on lessons learned with radio air shower arrays that have first demonstrated the experimental power of the radio detection technique, e.g.~\cite{Aab:2016eeq,Schellart:2013bba}.

As outlined above, a location is needed with thick, homogeneous and cold ice to yield the best experimental results. An additional requirement is the availability of a sufficiently developed infrastructure to allow for installation, running and maintenance of the detector. While the instrumented stations can be fully autonomous, the amount of cargo and personnel needed for installation requires accessibility by plane or large vehicle. 
The number of accessible research stations fitting these requirements in either Antarctica or Greenland is limited.
The host institutions of the RNO-G collaboration members and their access to national infrastructure additionally excludes some obvious candidate sites (Dome A, Dome C and Vostok in Antarctica, e.g.), leaving essentially South Pole Station and Summit Station in Greenland. South Pole station already houses a premier CMB instrument (the South Pole Telescope \cite{2011PASP..123..568C}), as well as the world's largest neutrino telescope (IceCube), which is in the process of installing the IceCube-Upgrade \cite{Ishihara:2019aao}. The
logistical burden is, thus, already high at South Pole.

If RNO-G is also to be used to develop and test hardware for the radio component of IceCube-Gen2 \cite{Aartsen:2020fgd}, a site similar to South Pole has advantages, if South Pole station is unavailable. Interesting coastal sites, like the Ross-Ice-Shelf close to McMurdo Station, which hosts the ARIANNA experiment \cite{ariannaWhitepaper}, can assist in developing other technologies, but would be unable to replicate some of the particular challenges of South Pole. 

To achieve a high trigger efficiency, a cosmic-ray veto, and the ability to reconstruct events with high accuracy, the RNO-G design combines a surface with a deep array capable of operating at low threshold (see Fig.~\ref{fig:layout}). The collaboration will develop the necessary expertise for rapid installation with a minimum of logistical impact, enabled by newer, fast drilling technology and lightweight, low-power, autonomous stations that still achieve excellent single-station effective volume.

\subsection{Summit Station, Greenland}
\label{sec:summit}
Going to Greenland also has some fundamental consequences for the design decisions. 
The Antarctic has been host to several pioneering arrays that aim to detect in-ice radio emission from UHE neutrinos. Through previous efforts, the Arctic has been established as a parallel site for a future radio neutrino observatory~\cite{avva1, Avva:2016ggs, Deaconu:2018bkf}. Summit Station offers several advantages as a testbed site. It is located at $72^{\circ}35'46''$~N, $38^{\circ}25'19''$~W at the peak of the Greenland ice cap, atop more than 3~km of glacial ice that we have measured to be remarkably radio transparent~\cite{avva1} at $\sim$100 MHz, and with a $\sim$100~m deep firn layer that we have preliminarily characterized~\cite{Deaconu:2018bkf}.  It is a year-round scientific research station sponsored by the National Science Foundation. It has a snow runway that accommodates LC-130 Hercules flights to deliver cargo and personnel, and facilities on site to support science. Compared to sites in Antarctica, Summit Station ($72^\circ$~N Latitude) is easier to access from the Northern hemisphere, in particular through commercial flights from Europe, and has a larger fraction of the year with daily periods of light, providing a higher livetime for autonomous solar-powered stations. This final aspect is particularly important, given the reduced electrical generator infrastructure at Summit compared to South Pole. The restriction to renewable energies, combined with battery buffering limitations and the desire for high livetime, cap the amount of power the detector can draw and ultimately drive the station design.

Logistical considerations at Summit also favor a compact geometry with fewer, more sensitive stations rather than more, less sensitive stations. Similarly, the drilling technique must be light-weight and mobile and, therefore, mechanical.

The ASIG drill, which is able to drill \SI{5.75}{''} diameter boreholes to \SI{100}{m} at a rate of 1 hole per day, was initially considered as the main option \cite{ASIG}; subsequent antenna design was adapted to that form factor. Alternatively, the British Antarctic Survey (BAS) has been developing a mechanical drill that provides larger boreholes of \SI{11.2}{''}, which will allow for greater flexibility in antenna design. Both drills satisfy the drilling rate, hole diameter and logistical impact specifications. See Sec.~\ref{sec:deployment} for an in-depth discussion of drilling and installation. 

To compensate for the warmer, more attenuating ice in Greenland compared to South Pole, triggering is performed with the deeper antennas, below the firn.
Since no detector has detected the radio emission following a neutrino interaction yet, the exact experimental signature is predicted by simulations only, arguing for a detector design that detects the neutrino signal in a multitude of channels to increase confidence. It can be considered to adapt and simplify the detection strategy once the first neutrino has been conclusively identified.

\subsection{A low-power, low-threshold trigger and data acquisition system}
\label{sec:low_threshold}

The RNO-G stations are built around an interferometric phased array, similar to what has been demonstrated {\it in situ} at the South Pole on the ARA experiment~\cite{oberla, Avva:2016ggs}, achieving the lowest sustainable signal trigger threshold demonstrated in the field. Since the astrophysical neutrino flux shows a falling spectrum, improved sensitivity to lower-energy events dramatically increases the detected neutrino event rate.  The phased-array technology has been adapted for RNO-G to provide similar sensitivity, albeit with a reduced power consumption. 

\begin{figure}
\includegraphics[trim={0 0.6cm 0 0},clip,width=\textwidth]{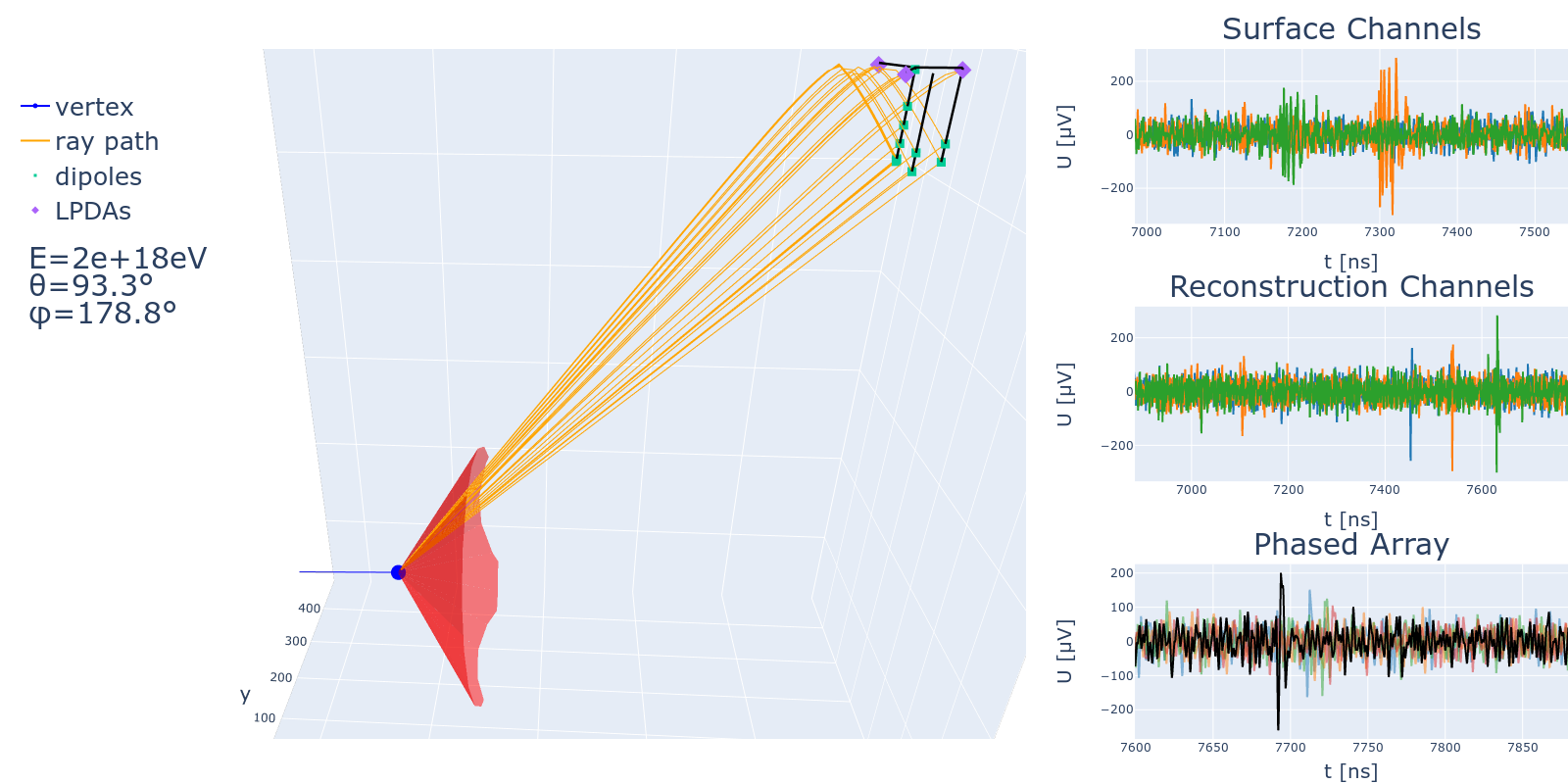}
\caption{A simulated RNO-G neutrino event. The left side shows the event geometry illustrating both the direct and reflected ray-paths to the antennas, as well as the incoming neutrino and interaction vertex (blue) and its Cherenkov cone (red), where the strongest signals are expected. The right side shows the waveforms in selected antennas, and the improvement in signal-to-noise-ratio obtained by phasing the signals as done in the phased array trigger (shown in black, bottom row). This simulation and online event display utilize tools developed by the greater radio community \cite{NuRadioReco,NuRadioMC}. For better visibility, only selected channels are shown.
} 
\label{fig:display}
\end{figure}

The phased array trigger coherently sums single channel waveforms with time delays corresponding to a range of angles of incident plane waves, improving the trigger-level signal-to-noise ratio roughly linearly with the number of antennas in the array \cite{oberla}, as illustrated in Fig.~\ref{fig:display}.
Projecting the performance of the existing ARA system, we expect to achieve an elevation-averaged 50\% trigger efficiency point at a 2$\sigma_\textrm{noise}$ threshold in voltage. This low threshold is needed to observe the largest volume of ice possible as discussed in Sec.\ref{sec:emission}. 

It should be noted that for the simulations, the following definition for signal-to-noise ratio (SNR) and noise is handled. SNR is defined as the amplitude of the noiseless signal over the standard deviation $\sigma_\textrm{noise}$ of a pure noise waveform. A threshold of 2$\sigma_\textrm{noise}$ thus means a threshold of twice the standard deviation of a trace without signal.

\subsection{Detector geometry: An integrated approach with deep and surface components}

\begin{figure}
\centering
\includegraphics[width=0.6\textwidth]{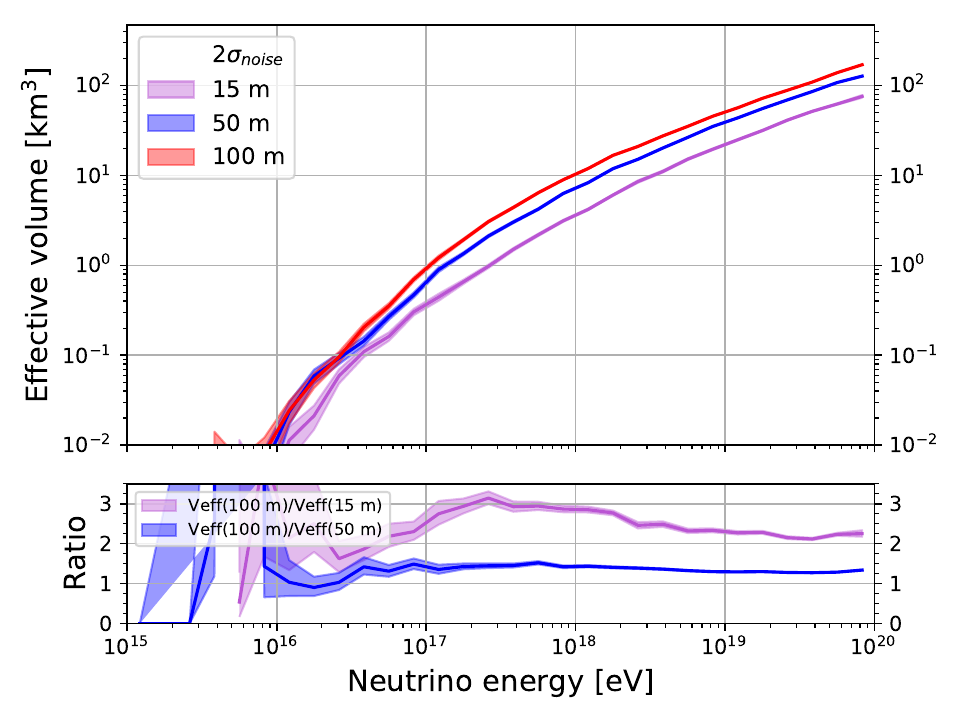}
\caption{Scaling of the effective volume as a function of the depth of antennas used for the phased-array trigger for 35 stations deployed at Summit Station. \SI{100}{m} was used during the design process as the technological limit due to drilling restrictions. As the trigger antennas are placed deeper into the ice, the effective volume increases, due to the number of allowed ray trajectories in the ice.
The curves have been obtained using a $2\sigma_\textrm{noise}$ dipole proxy for the phased array. Shown are statistical uncertainties only. Below \SI{10}{PeV} the uncertainties on the effective volume become too large to draw firm conclusions. 
}
\label{fig:analysis_eff_volume}
\end{figure}

After extensive trade studies, we have coalesced on a station design that integrates a deep component with a surface component, as shown in Fig.~\ref{fig:layout}.  This integrated design achieves the highest effective volume per station given the phased-array trigger and mechanical drilling technology to \SI{100}{m}. As shown in Fig.~\ref{fig:analysis_eff_volume}, the effective volume per station increases with increasing depth, so our design places the deep component of the station as deep as is logistically feasible, given the current constraints of drilling.

In addition to maximizing effective volume, the station design has been optimized for neutrino reconstruction efficiency. The string containing the phased array trigger will feature additional Vpol antennas almost equally spaced vertically along the string allowing us to pin-point the neutrino vertex and zenith angle of the signal arrival direction, and achieving high accuracy by exploiting azimuthal symmetry. Since the down-hole Vpol antennas are typically more sensitive than the Hpol antennas and the trigger selects signals having a measurable component in the vertical polarization, the Vpol antennas dominate vertex and signal arrival direction reconstruction. Adding two Hpol antennas above the phased array will allow us to improve the reconstruction of the full electric field. Combining the four Vpol antennas of the phased array with the two Hpols in proximity, should provide sufficient information to reconstruct the polarization of the signal, as well as its frequency slope, and thereby the off-Cherenkov signal angle and neutrino arrival direction.

The radio signal from a neutrino interaction often travels along direct, refracted or reflected paths (designated \textit{DnR}) to the deep array, as shown in Fig.~\ref{fig:display}. The characteristic double pulse would be a smoking-gun signature of an in-ice source. The difference in direct and refracted arrival times significantly improves the reconstruction of the neutrino vertex position, and thereby the shower energy, as well as arrival direction \cite{Anker:2019zcx}.  The probability to observe both a direct and a reflected signal is depth dependent. The spacing of the Vpol antennas on the main string is the result of an optimization between double pulse detection and long lever arm for good angular reconstruction.

The two additional deep boreholes are needed for a full direction reconstruction. Three independent measurements are needed for azimuthal information, which is provided by the Vpol antennas. By placing the Hpol antennas at different depths on every string, both zenith and azimuth information will be provided for those signals with a strong horizontal polarization component, as well as increasing the probability to reconstruct a signal for those events with little signal strength in the horizontal component. 

The additional strings also host the calibration pulsers, which will ensure regular monitoring of the performance of the station and provide information useful for precise calibration of the antenna geometry.  In addition, a surface pulser is foreseen, which will be deployed in a hand-drilled hole below the surface.

The surface component will deliver precision polarization measurements and timing information for all events detected at the surface. Also, the broad-band sensitivity of the log-periodic dipole antennas (LPDAs) will broaden the frequency coverage of the detector, which helps determine the radio detection angle with respect to the Cherenkov cone, improving energy reconstruction and pointing resolution. Events detected only in the surface components, however, only add minimally to the total neutrino effective volume.

With the planned layout, any events observed in coincidence between the surface component and the deep component are particularly valuable for event reconstruction; the fraction of these events is discussed in Sect.~\ref{sec:low_bg}. In addition, the surface channels serve as an efficient air shower veto, reducing the background for neutrino searches as will be discussed in the following section. 

\begin{figure}
    \centering
    \includegraphics[width=.5\columnwidth]{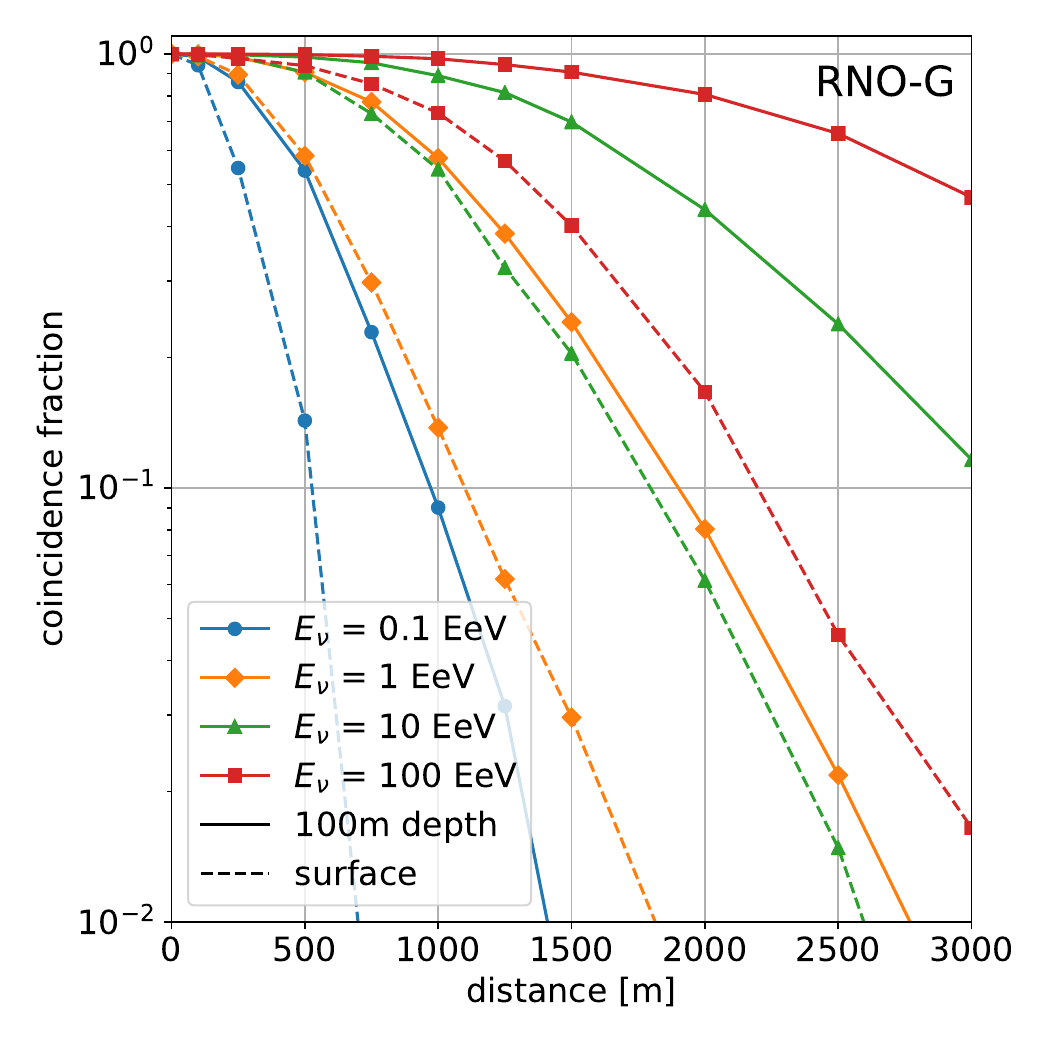}
    \caption{Fraction of coincident triggers on nearby stations as a function of the grid distance between stations for different neutrino energies. In addition to the \SI{100}{m} deep power string, the coincidence fraction expected for a trigger near the surface is given for comparison.}
    \label{fig:coincident_triggers_stations}
\end{figure}

The stations will be deployed on a square grid with \SI{1}{km} baseline. This means that at energies beyond \SI{1e18}{eV} the effective volumes of the stations start to overlap and coincident measurements of the same neutrino become likely. This can be seen from Fig.~\ref{fig:coincident_triggers_stations}, where the fraction of events triggered in coincidence is shown for different neutrino energies and grid spacings. While limiting the total effective volume of the system, \SI{1}{km} was chosen to restrict the logistical impact in installation and preserve the opportunity of coincident events, which will simplify event identification and provide excellent reconstructed properties. As the project advanced, one may consider spacing stations further apart. 

\subsection{High analysis efficiency and low background to enhance discovery potential} 
\label{sec:low_bg}

In addition to triggering on and extracting event parameters from neutrino events, we must be able to separate any neutrino events in our recorded data set with high efficiency from all backgrounds. The three major sources of background are incoherent thermal noise, impulsive anthropogenic noise, and radio impulses resulting from cosmic-ray air showers.  
A discovery experiment of this scale requires low backgrounds at the level of 0.01 per station per year (or less). RNO-G is designed to achieve this ambitious background level by building on two key measures that have been developed to ensure event purity.

\begin{itemize}
    \item[(1)] Triggering from deep in the ice (at a depth of \SI{100}{m}), where the backgrounds are smaller than at the surface: ARA has shown that the anthropogenic and thermal backgrounds decrease for receivers deployed deeper in the ice \cite{Allison:2014kha} and further from human activity at research stations, achieving a background on the most recent analysis of 0.01 events in two stations over 1100 days of livetime~\cite{ara_2station}. This shows that successful background rejection can be achieved also during the summer at South Pole when anthropogenic backgrounds are more significant.  
    \item[(2)] Vetoing backgrounds using a surface detector component: Non-thermal backgrounds are introduced from the surface or from close to the surface by man-made sources or air shower remnants. Surface antennas will help to separate neutrino induced signals originating within the ice from those of air showers \cite{deVries} and those from showers caused by catastrophic energy losses from atmospheric muons \cite{Garcia-Fernandez:2020dhb}. 
\end{itemize}  

\begin{figure}
    \centering
    \vspace{-\baselineskip}
    \includegraphics[width=.98\textwidth]{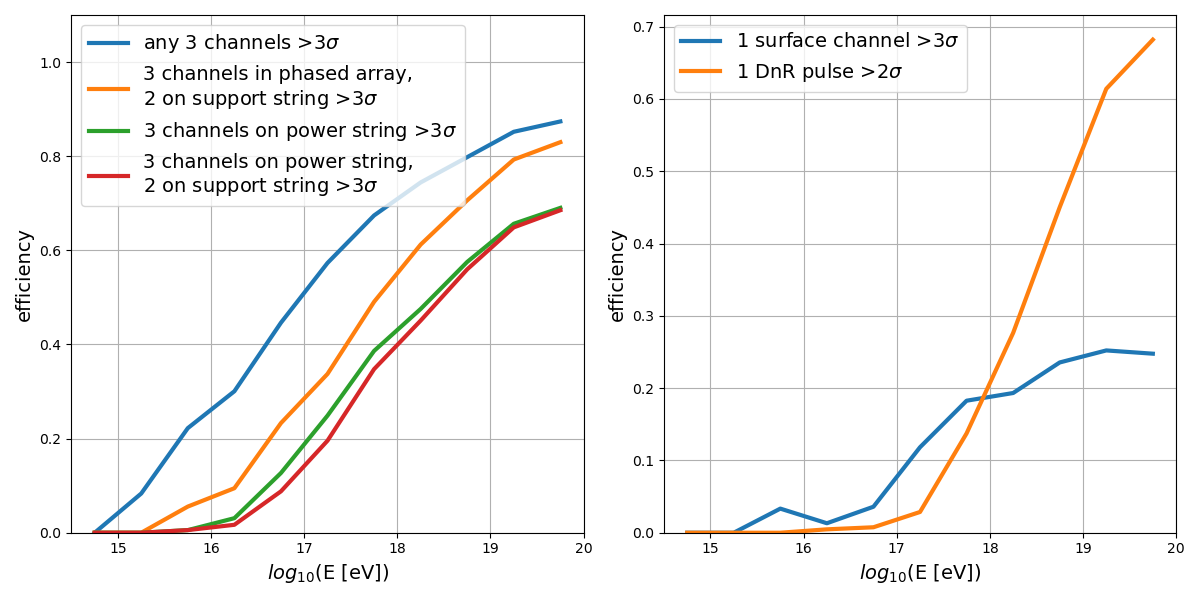}
    \caption{Estimate of anticipated analysis efficiency, defined here as the fraction of events recorded in a pre-defined number of antennas (channels) above a given threshold. The left figure illustrates detections in the deep part of the array only. 
    A detection in the support string (see Fig.~\ref{fig:layout}) will allow for a reconstruction of the arrival direction of the signal and three antennas on the power string are needed for a vertex reconstruction. 
    The right figure shows the fraction of DnR signals, which are particularly valuable for the vertex reconstruction and the fraction of very valuable events measured in both the deep array and the surface antennas.
    For these simulations we assume a trigger at 2.0$\sigma_\textrm{noise}$ trigger in the phased array.  
    }
    \label{fig:analysis_efficiencies}
\end{figure}

Neutrino events triggered near threshold in the phased array system carry the risk to have low SNR in antennas needed for reconstruction. The information content in different numbers of antennas is illustrated in Fig.~\ref{fig:analysis_efficiencies}. Three channels detecting a signal $>$3$\sigma_\textrm{noise}$ is taken as a simple proxy for events that can be identified and reconstructed with currently available analysis techniques such as interferometry~\cite{Romero-Wolf:2014pua}, template matching~\cite{arianna15a} and signal de-dispersion~\cite{anita3}.
With a signal in the antennas of the phased array as well as in an antenna on the support string, it is possible to reconstruct the neutrino arrival direction (see Sect.~\ref{sec:angular_sensitivity}). To reconstruct the shower energy, at least 3 of the reconstruction antennas on the \emph{power string} need to detect a pulse so that the distance to the interaction vertex can be reconstructed (see Sect.~\ref{sec:energy}). In some cases, the radio signal is reflected off the ice-air interface or diffracted downwards, so that two signals from the same shower can be detected. These so-called \emph{DnR pulses} become more likely with higher neutrino energies and can be used to greatly improve the reconstruction accuracy. More details on reconstruction and resolution is given in Sect.~\ref{sec:energy} and \ref{sec:angular_sensitivity}. Foreseeable advances in analysis techniques will further improve the efficiency near threshold, both in firmware and also in off-line analysis.

\begin{figure}
    \centering
    \includegraphics[width=0.49\textwidth]{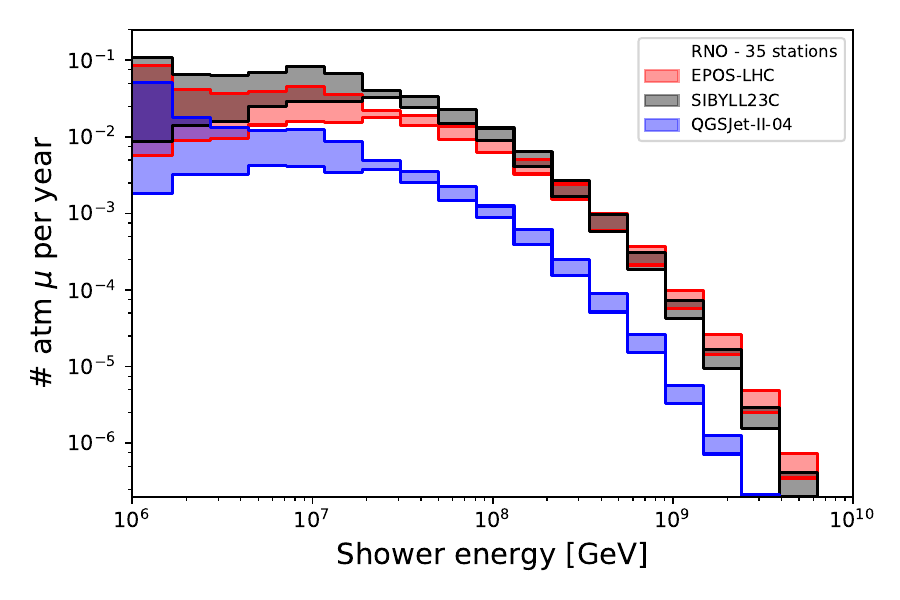}
    \includegraphics[width=0.49\textwidth]{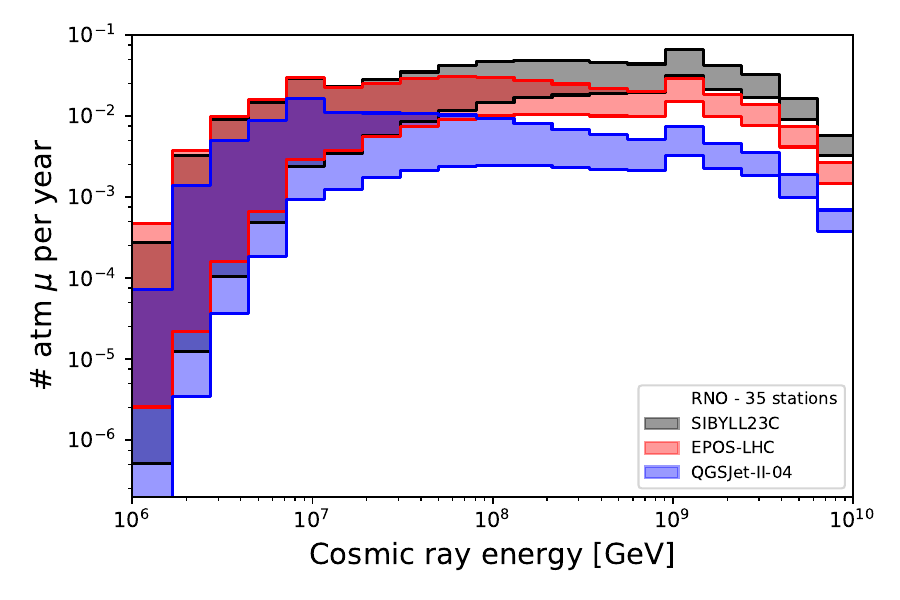}
    \caption{Number of atmospheric muons detected by a 35-station array at Summit Station. The phased array is modeled with dipoles having amplitude thresholds varying from $1.5$ to $2.5\sigma_\textrm{noise}$, at
    \SI{100}{m} of depth. Each color represents a different hadronic model, as specified in the legend. The bands include the range of expected events for the different simulated thresholds as well as the 68\% CL contour corresponding to the effective area uncertainty.
    Left: number of detected atmospheric muons per year as a function of shower energy. Right: same results, presented as a function of cosmic ray energy. The drop off at low energies is an artifact of only simulating muons down to \SI{1e6}{GeV}.
    }
    \label{fig:muon_background}
\end{figure}

An accurate knowledge of the existing background is needed in order to project what fraction of triggers are due to non-neutrino backgrounds, and also to assess whether a veto mechanism is advisable (or mandatory). An important type of background which is difficult to distinguish from actual neutrino events is the background created by the energy losses from atmospheric muons from cosmic ray showers. These muons are produced in the atmosphere, continue their propagation into the ice; their subsequent interactions (mainly bremsstrahlung, photonuclear interaction, and pair production) create hadronic and electromagnetic showers that emit radiation and are therefore detectable by an in-ice radio array \cite{Dutta:2000hh}. 
These muons share shower characteristics, arrival directions and vertex positions with the sought-after neutrinos (see \cite{Garcia-Fernandez:2020dhb}).
We have calculated the number of expected muon-initiated showers for a 35-station array at Summit Station, using a \SI{100}{m}-deep dipole with an amplitude threshold between $1.5\sigma_\textrm{noise}$ and $2.5\sigma_\textrm{noise}$ as a proxy for the phased array.
The effective areas have been calculated using NuRadioMC \cite{NuRadioMC} and its interface to PROPOSAL \cite{proposal}. Then, these effective areas are convolved with the expected muon flux at the detector, calculated by MCEq \cite{mceq}.  The chosen cosmic ray flux model is the Global Spline Fit from \cite{GSF}. This procedure is explained more in detail in \cite{Garcia-Fernandez:2020dhb}. The results are presented in Fig.~\ref{fig:muon_background}, where each band represents the results for a hadronic interaction model. Shown are the expected number of detected muons for the phased array proxies (from $1.5$ to $2.5\sigma_\textrm{noise}$) and also the 68\% CL interval for the uncertainty due to cosmic ray flux, hadronic modeling and effective area. Fig.~\ref{fig:muon_background}, left, contains the expected detected number of muons per year for 35 stations as a function of shower energy, while Fig.~\ref{fig:muon_background}, right, presents the same results as a function of cosmic ray energy.

The lower and upper bounds on the number of detected atmospheric muons per year for a 35-station layout, as well as the average number for a $2.0\sigma_\textrm{noise}$ trigger, can be found in Table~\ref{tab:muon_numbers}. While these values are relatively benign, we remind that the neutrino event rate may be equally low, at least at the threshold energy of the detector. Hence, the air shower self-vetoing on the detector deserves special attention, as well as the development of algorithms using event parameters such as arrival direction and vertex location to disentangle neutrino signals from those potential background events. Also, since the flux and composition of cosmic rays at the relevant energies is subject to large uncertainties, those same uncertainties propagate into the background prediction for radio arrays. 

\begin{table}
\begin{center}
\begin{tabular}{ |c|c|c|c| } 
 \hline
                 & $2.5\sigma_\textrm{noise}$ 68\% CL LB & $2.0\sigma_\textrm{noise}$ average &  $1.5\sigma_\textrm{noise}$ 68\% CL UB \\ \hline \hline
SIBYLL 2.3C     & \SI{0.212}{} & \SI{0.296}{} & \SI{0.684}{} \\ \hline
EPOS-LHC        & \SI{0.129}{} & \SI{0.173}{} & \SI{0.444}{} \\ \hline
QGSJet-II-04    & \SI{0.031}{} & \SI{0.044}{} & \SI{0.180}{} \\ \hline
\end{tabular}
\caption{Number of detected atmospheric muons per year
for a 35-station layout. Three hadronic models
are shown. The numbers shown are the lower 68\% CL lower bound for
a $2.5\sigma_\textrm{noise}$ trigger (first column), the average values for a $2.0\sigma_\textrm{noise}$ trigger, and the 68\% upper bound for a $1.5\sigma_\textrm{noise}$ trigger.
See text for details.}
\label{tab:muon_numbers}
\end{center}
\end{table}

\section{The RNO-G instrument design}
\begin{figure} 
\includegraphics[trim={0 0.5cm 0 0.5cm},clip,width=\textwidth]{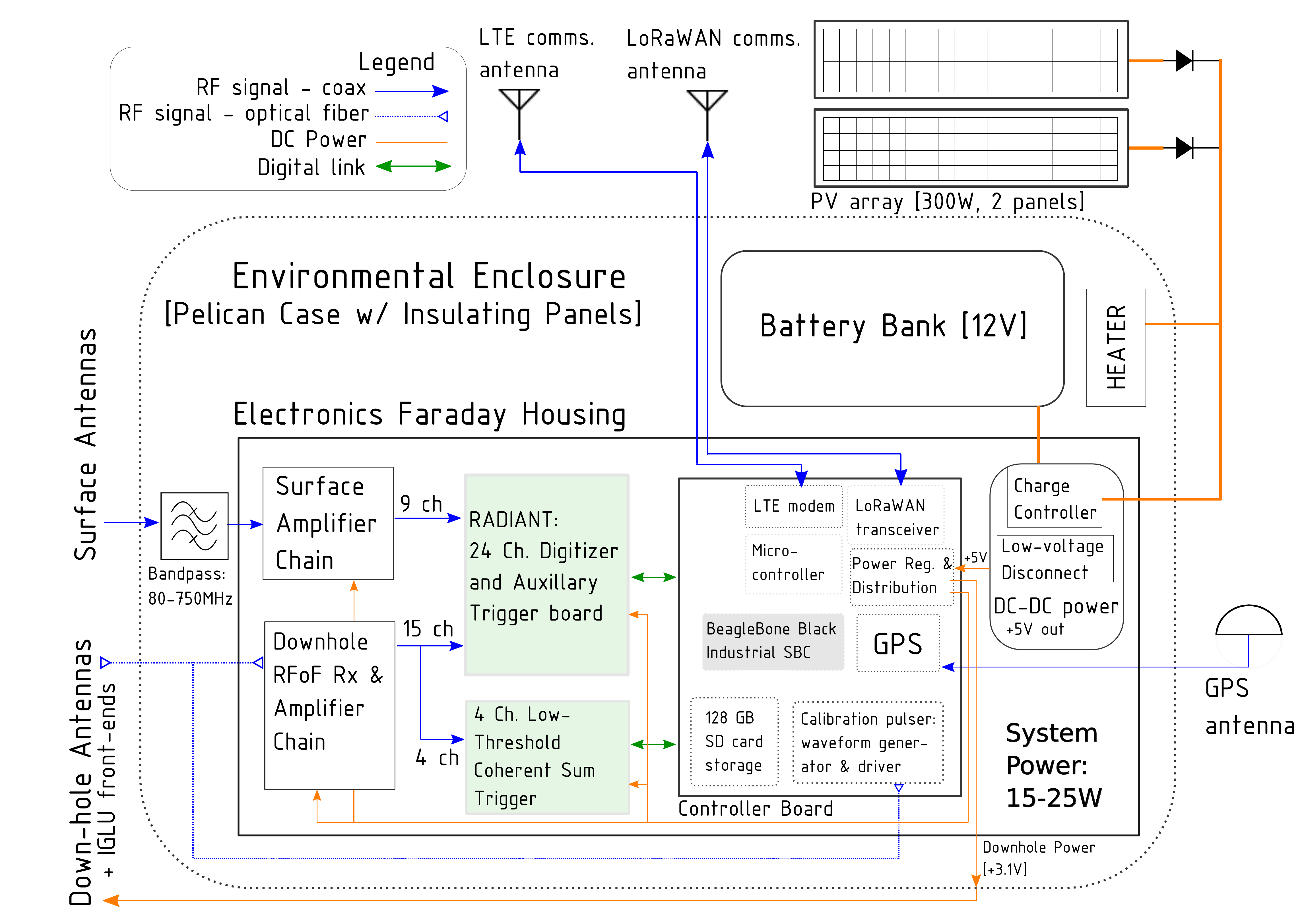}
\caption{System diagram for an RNO-G station. See text for details. 
}
\label{fig:system_diagram} 
\end{figure}

RNO-G will provide high-quality science data and a robust, low trigger threshold with minimal power consumption using a station design schematically depicted in Fig.~\ref{fig:system_diagram}. In nominal operating mode, a station will use \SI{25}{W}, including DC-DC converter losses.
All equipment is rated to operate at $-40^{\circ}\text{~C}$ and \SI{3200}{m} altitude. 

\subsection{Antennas}

\begin{figure}
    \centering
    \includegraphics[width=0.6\textwidth]{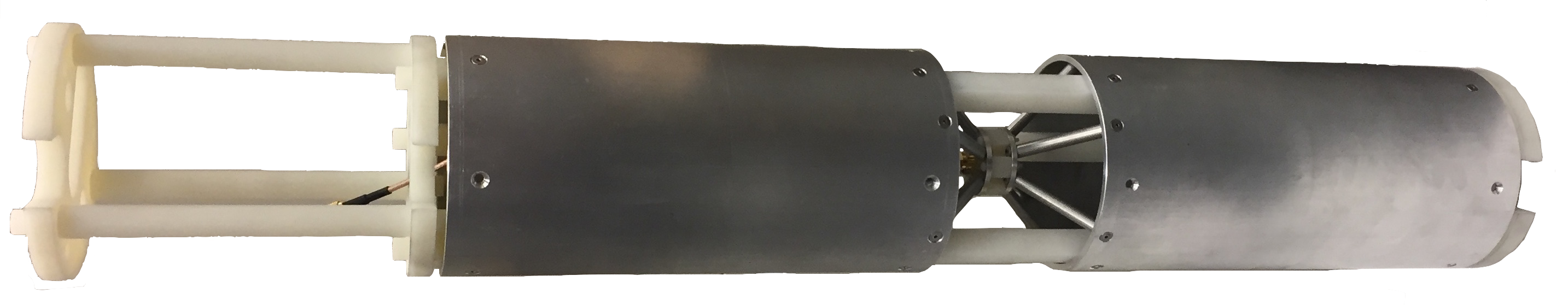}
    \includegraphics[width=0.6\textwidth]{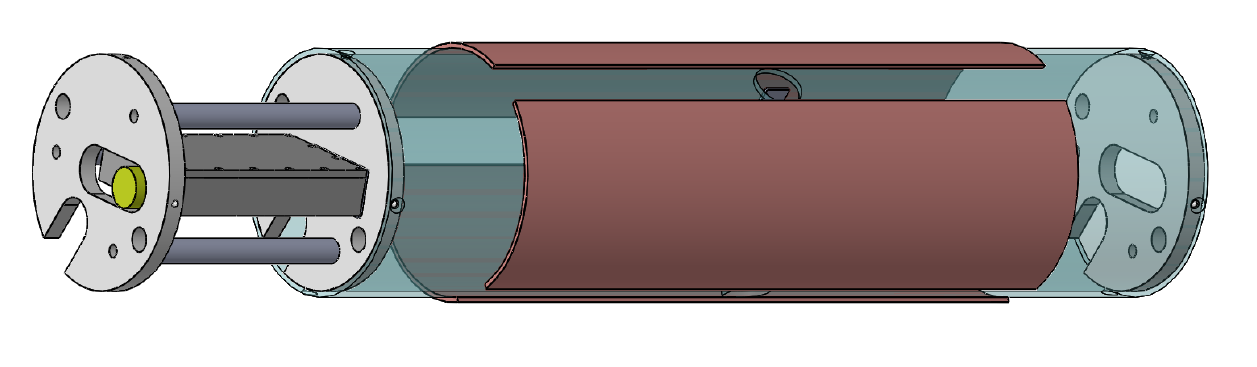}
    \includegraphics[width=0.22\textwidth,angle=90,trim={10.0cm 0.1cm 10.0cm 0.1cm},clip=true]{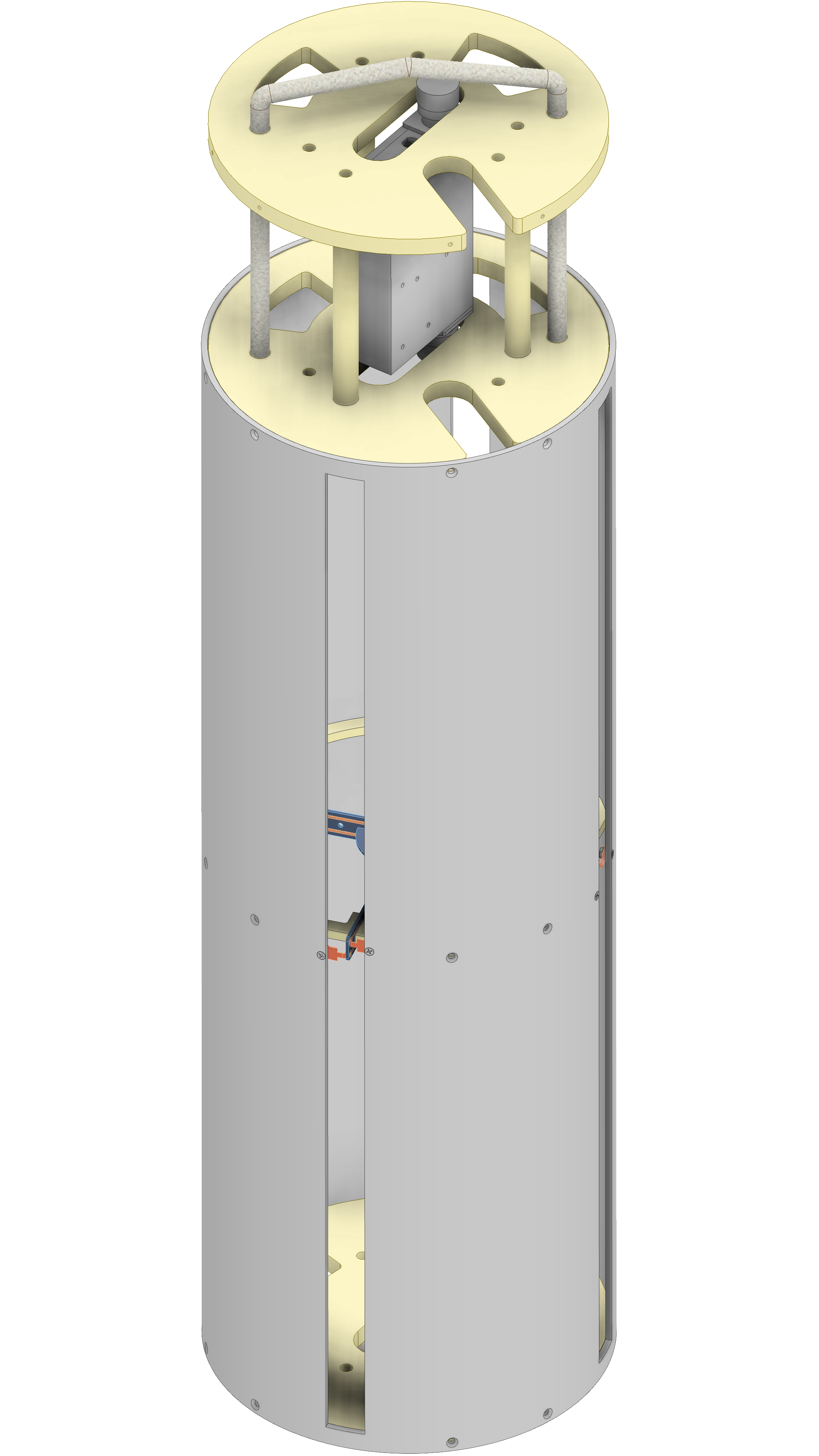} 
    \caption{Photo of a Vpol prototype (top) and technical drawings of options for the Hpol antennas (trislot, middle, quadslot, bottom). The Vpol and trislot are the first iterations of the deep antennas for RNO-G, while the quadslot is being considered for use in conjunction with larger diameter boreholes.
    }
    \label{fig:antennas}
\end{figure}

\begin{figure}
    \centering
    \includegraphics[width=100mm]{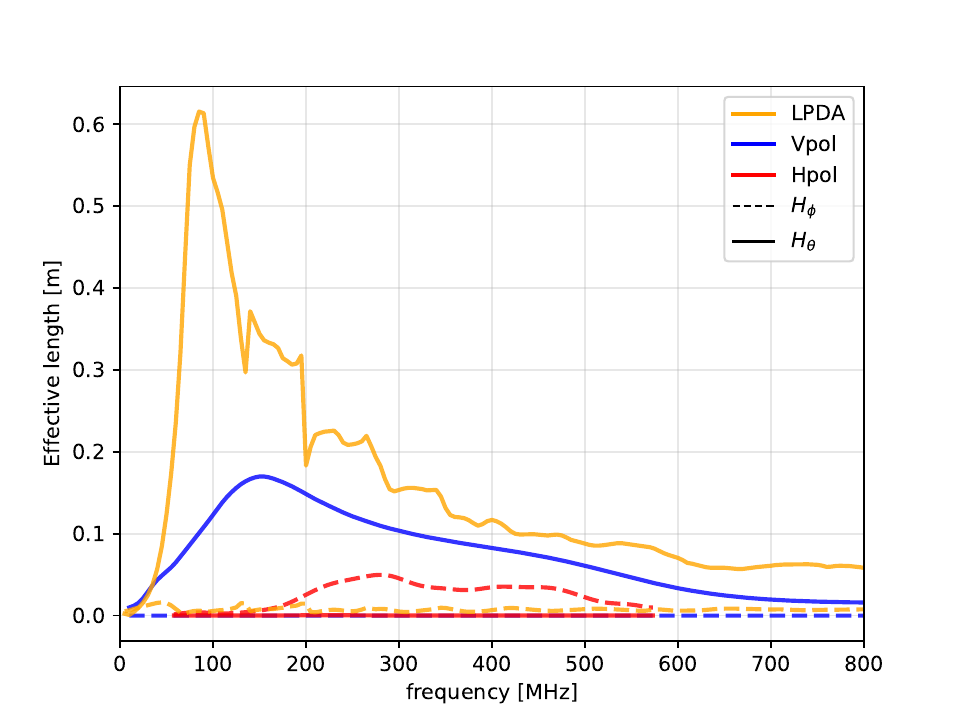}
    \caption{Antenna effective length magnitude for the LPDA, Vpol and Hpol (tri-slot) in the direction of maximum gain $H_{\theta}$ (V-pol and LPDA) or $H_{\phi}$ (H-pol). Results of detailed antenna simulations of the v1 iterations as shown in Fig.\ref{fig:antennas} 
    }
    \label{fig:antennapattern}
\end{figure}

The initial downhole antenna designs are driven by the \SI{5.75}{''} diameter of the boreholes (ASIG drill \cite{ASIG}), with some modifications possible, if bigger boreholes are available (see Sect.~\ref{sec:summit}. 
The vertically-polarized (Vpol) antennas will be a \textit{fat dipole} design (see Fig.~\ref{fig:antennas}) previously used in neutrino detection experiments, which have an azimuthally symmetric beam pattern and usable bandwidth ranging from 150-600 MHz \cite{rice03,Avva:2016ggs}. For horizontal polarization (Hpol), cylindrical \textit{tri-slot antennas} are considered. They are nearly azimuthally-symmetric in gain, with differences of less than 1~dB up to \SI{800}{MHz}, which corresponds to differences of less than 12\% in effective length. Only Vpol antennas are used for the trigger because the Hpol antennas inherently have narrower usable bandwidth than the fat dipoles, as shown in Fig.~\ref{fig:antennapattern}. With the current Hpol designs, there is enough overlap with the Vpol band to combine the signals for polarization reconstruction in analysis. Larger boreholes (RAID drill) will especially help improve the broadband characteristics of the Hpol antennas. It is under consideration to exchange the tri-slot design for 8'' quad-slot antennas, which will have a lower frequency turn-on and improved gain characteristics taking advantage of the larger allowed diameter. 

The surface component employs commercially available log-periodic dipole antennas (LPDAs, Create CLP-5130-2N), successfully used by the ARIANNA experiment. ARIANNA's extensive in-field experience with these antennas will significantly simplify calibration. Owing to the high gain allowed without the borehole constraints, the nine LPDAs arranged in various orientations (see Fig.~\ref{fig:layout}, right) will measure all polarization components with high-precision, and provide a clear separation of upgoing versus downgoing signals. Due to their size the LPDAs have the largest gain of all employed antennas and will provide the greatest frequency coverage for the detected signals. 

Particular care will be taken to placement and alignment of the LPDAs in the trenches at the surface, as well as surveying the position of boreholes and antenna locations to ensure good starting values for the system calibration using the {\it in situ} pulsers.  

\subsection{Radio-Frequency front-end design}

\begin{figure}
\includegraphics[width=0.32\textwidth]{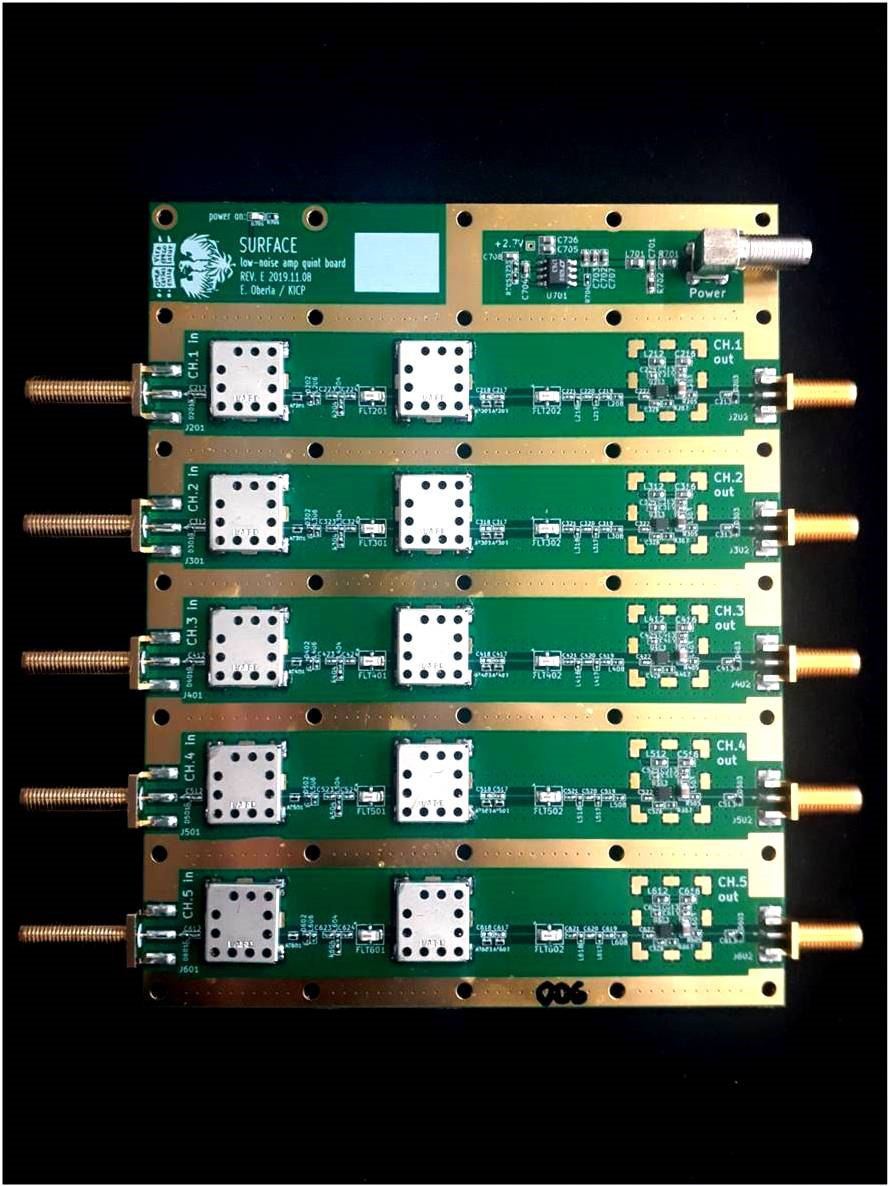}
\includegraphics[width=0.32\textwidth]{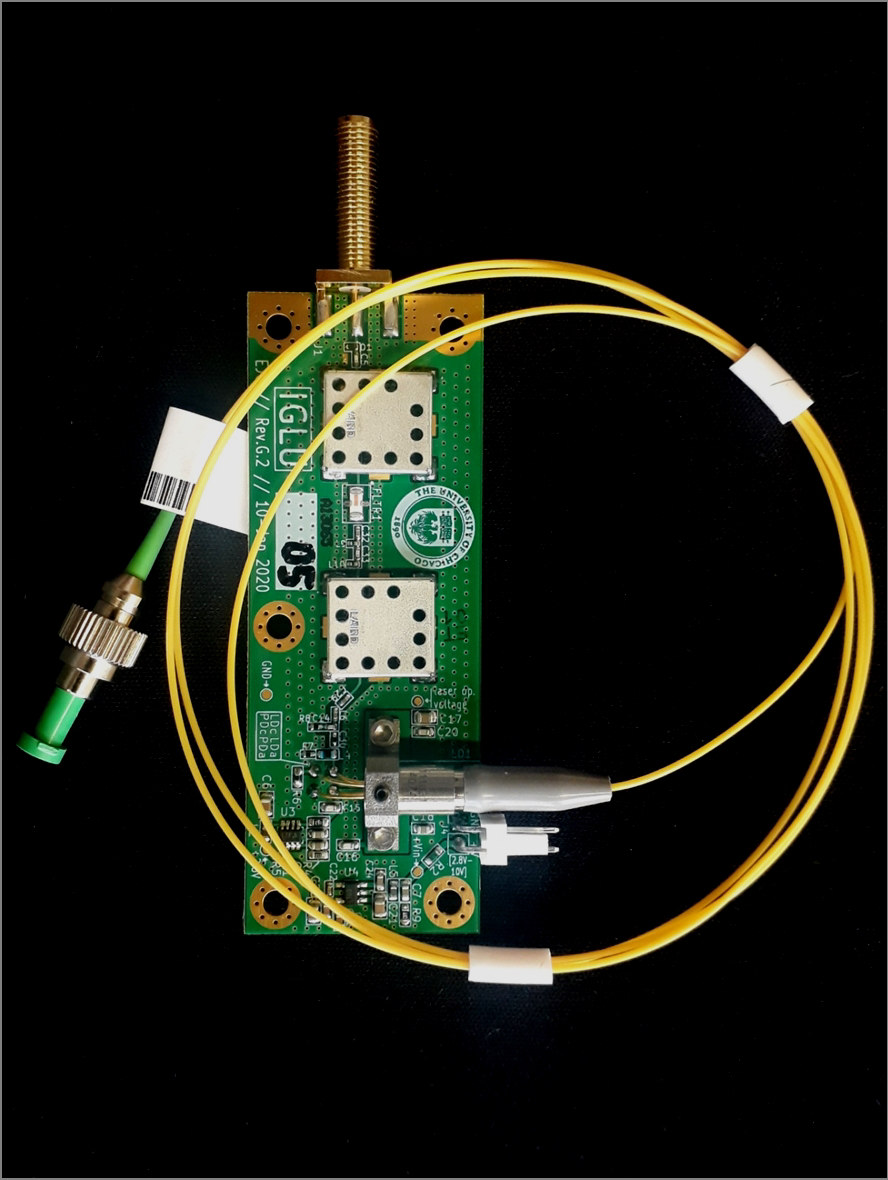}
\includegraphics[width=0.32\textwidth]{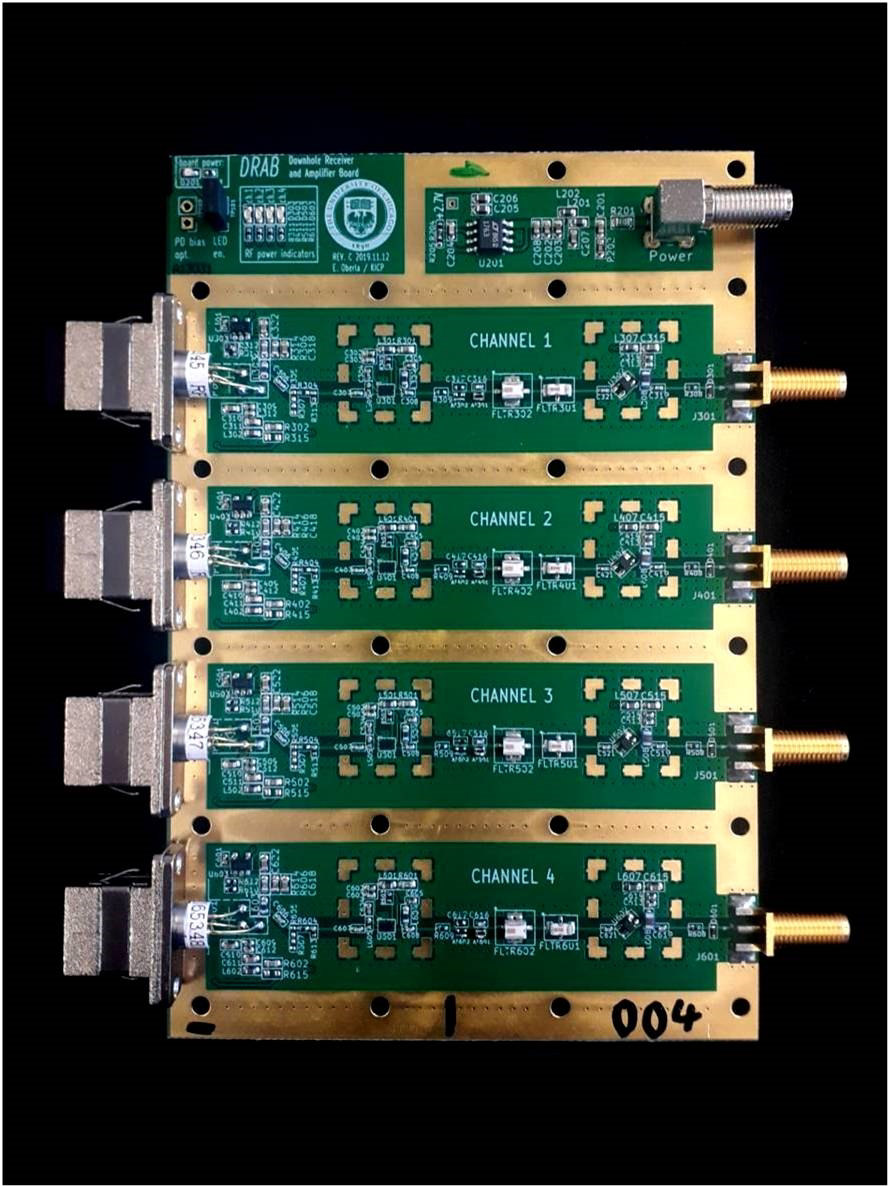}
\caption{Amplifiers as designed for RNO-G. Left: SURFACE amplifiers for the signals coming from the LPDAs via coaxial cable. Middle: an IGLU board (In-ice Gain with Low-power Unit) used to convert signals from antennas deep in the ice to analog RF signals and then feed them into the indicated fiber. Right: DRAB board (Down-hole Receiver and Amplifier Board) located within the station housing. All amplifiers are shown without their environmental enclosures. 
} 
\label{fig:photos_amps}
\end{figure}

To minimize system noise temperature, the feed of each antenna deployed in the borehole is connected with a short coaxial cable to a downhole front-end (Fig.~\ref{fig:system_diagram}, where a Low-Noise Amplifier (LNA, type IGLU, see Fig.~\ref{fig:photos_amps}) boosts the signal strength. To prevent a significant gain slope from long lengths of copper coaxial cable, each front-end contains a Radio Frequency over Fiber (RFoF) transmitter. The RFoF link and LNA are both powered by a DC connection from the surface, which is the only through-going coaxial cable in the boreholes. The LNA and RFoF are custom designs optimized for minimal noise temperature ($\leq$150~K) and low power. Each downhole channel consumes \SI{140}{mW}, compared to \SI{2.5}{W} in the previous installation of the phased-array in ARA. A total of 15 downhole antennas are distributed across three boreholes. 

After being transmitted over fiber, the signals are received by another set of amplifiers in the DAQ box (type DRAB, see Fig.~\ref{fig:photos_amps}) and converted back to analog signals. At the DAQ box, the signals from the surface channels are also received and amplified. Given the relatively short run of coaxial cable from the LPDAs to the DAQ box of less than \SI{20}{m}, the signals require only one amplification stage after being fed into the DAQ box (type SURFACE, see Fig.~\ref{fig:photos_amps}).

\begin{figure}
\includegraphics[width=0.49\textwidth]{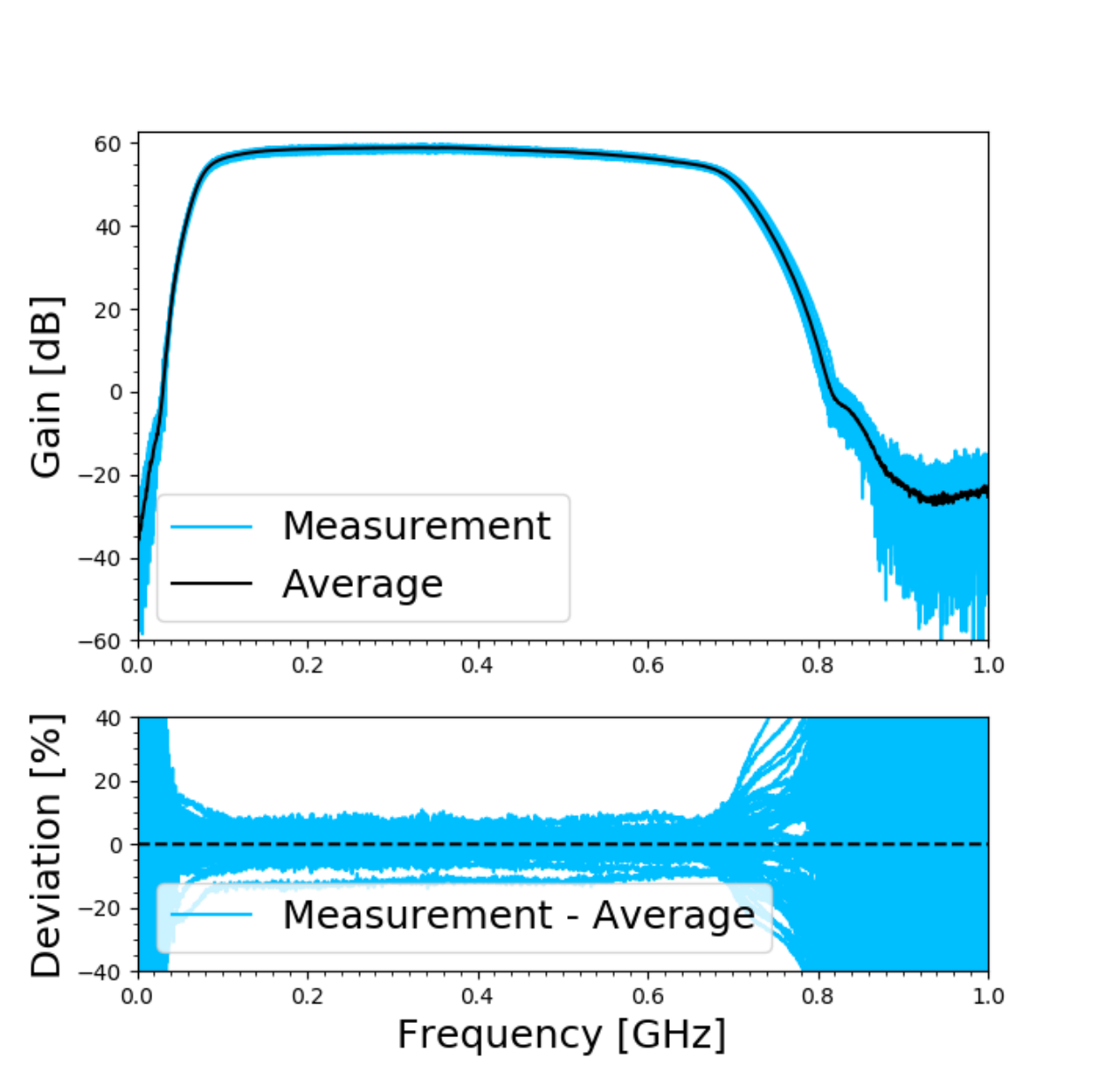}
\includegraphics[width=0.49\textwidth]{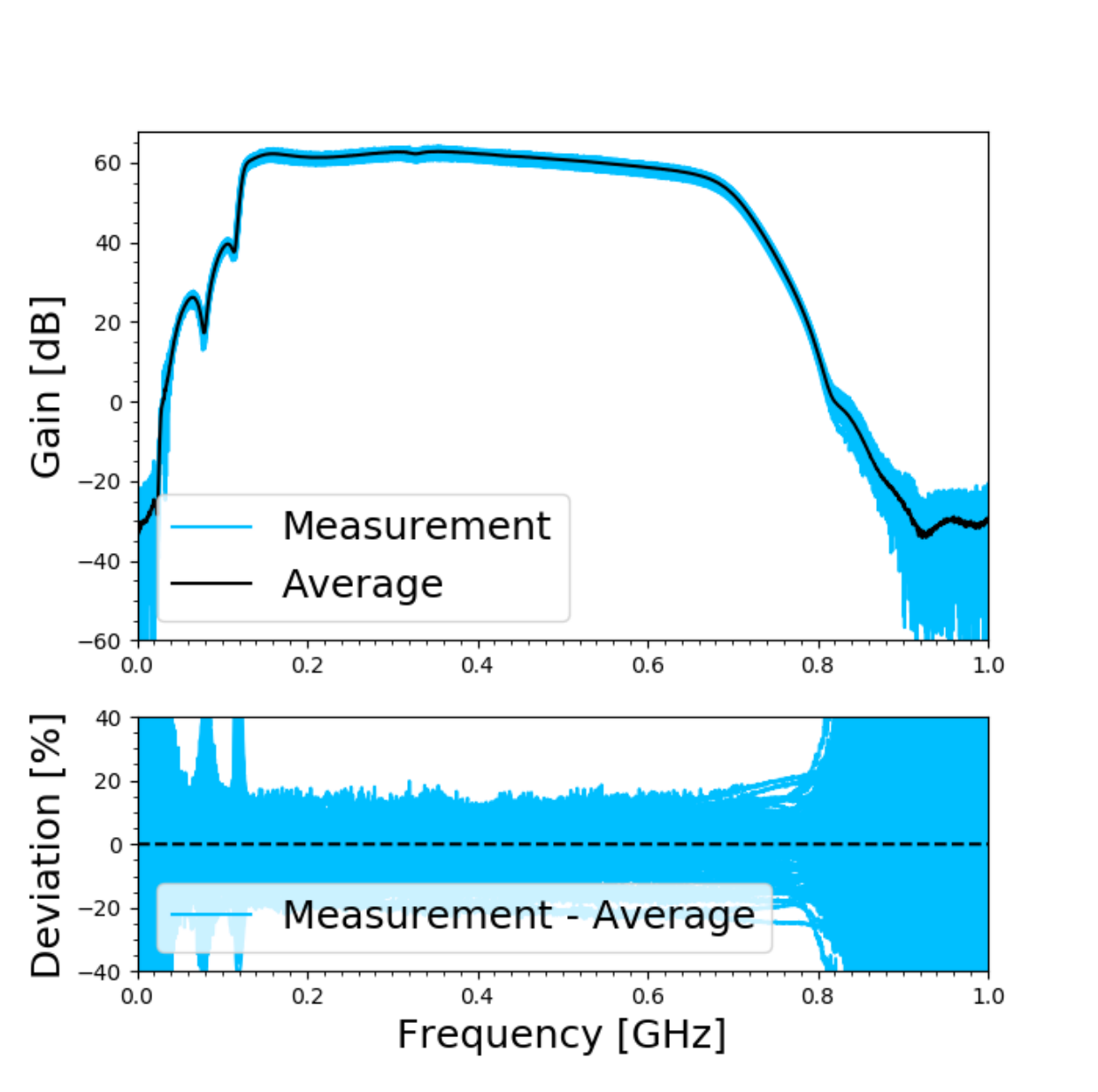}
\caption{Gain of the RNO-G amplifiers. Left: 12 SURFACE amplifiers. Right: Combination of 23 IGLU and DRAB amplifiers, including a \SI{50}{m} optical fiber cable. All amplifiers are revision v1 hardware.
} 
\label{fig:amp_gain_deviation}
\end{figure}

All amplifiers are placed in custom-designed RF-tight housings using iridited aluminium (chromate conversion coating). This significantly reduces the influence of noise on the amplifiers and protects the IGLU amplifiers in the boreholes from the environment. The amplifiers exhibit excellent uniformity in laboratory tests (see Fig.~\ref{fig:amp_gain_deviation}). Nevertheless, 
all amplifiers will be calibrated individually to reduce  systematic uncertainties on the reconstructed signals.

\subsection{Triggering, digitization, and data acquisition}
\label{sec:triggering}

\begin{figure}
    \centering
    \includegraphics[trim={0 0.7cm 0 0.7cm},clip,width=0.6\textwidth]{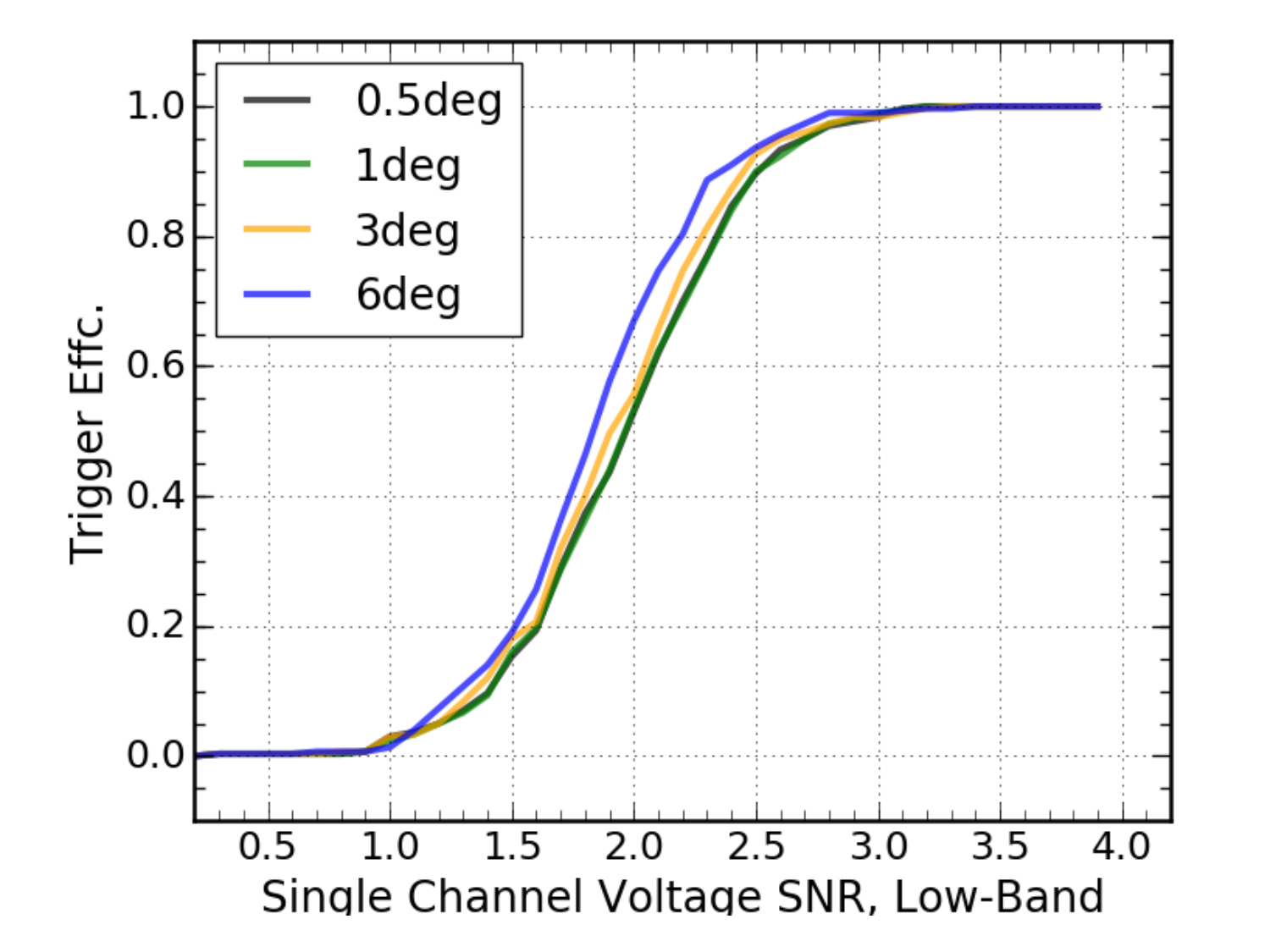}
    \caption{End-to-end simulation of the 4-antenna phased array trigger design for RNO-G. The simulated trigger efficiency for a number of neutrino signals at different off-cone viewing angles in the trigger bandwidth of \SIrange{80}{250}{MHz}.
    }
    \label{fig:phasedTrig}
\end{figure}

The main trigger of RNO-G will come from a phased-array at depth of \SI{100}{m}. The design of the field-proven phased-array installed at ARA \cite{oberla} had to be changed to accommodate the lower power requirements of autonomous stations and was optimized with respect to the neutrino signals typically expected in Greenland and with respect to per-item cost for the scalability of the array.  

The primary trigger will thus be a coherent-sum and beam-forming trigger from a compact array of four vertically-polarized antennas installed at the bottom of the main borehole string at a depth of \SI{100}{m}. A commercially available 8-bit \SI{500}{MSa/s} ADC is used to digitize and continuously stream data to an FPGA. This reduces the effective band to operate at the low-end of the signal bandwidth, \SIrange{80}{250}{MHz}. The lower cut-off is determined by the amplifier design that takes advantage of the full-range of low-frequency power that the antenna delivers. 

Eight beams will be formed that cover the full range of expected signal arrival directions. Compared to the previous phased-array implementation in ARA there will be fewer beams, but each of them wider, thus no angular coverage loss is incurred. Overall, the power-savings total to about a factor of 10 for the trigger board, using \SI{4}{W} in full operation mode. 

A single-antenna voltage threshold of 2$\sigma_\textrm{noise}$ can be achieved with this trigger, based on simulation studies as shown in Fig.~\ref{fig:phasedTrig}. The smaller bandwidth reduces the SNR of on-cone signals (i.e.\ 0.5 deg in Fig.~\ref{fig:phasedTrig}) by 10\%, however, increases the SNR for off-cone events by up to 80\%, thereby incurring very little loss on the absolute neutrino effective volume. This is due to the limited high-frequency content of off-cone neutrino signals (see also Fig.~\ref{fig:askaryan_pulses}). 

\begin{figure}
    \centering
    \includegraphics[width=5in]{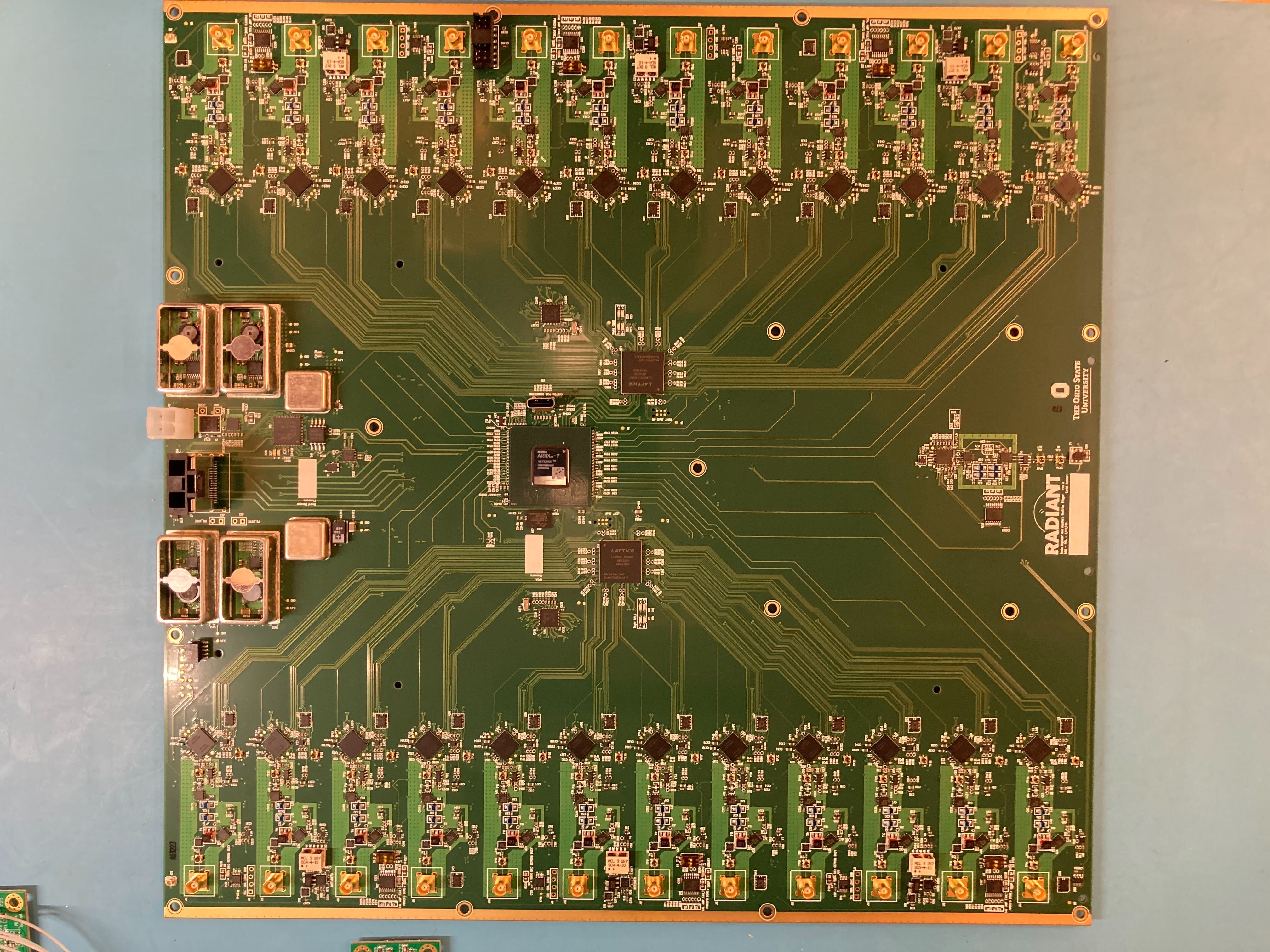}
    \caption{First iteration of the Radiant Board that will be the main DAQ of RNO-G. All 24 channels are accommodated on one board and read out by LAB-4D chips.  
    }
    \label{fig:DAQ}
\end{figure}

The full-band waveforms for all 24 antennas within a station will be digitized using the RAdio DIgitizer and Auxiliary Neutrino Trigger (RADIANT) board (Fig.~\ref{fig:DAQ}). The single-channel LAB4D switched-capacitor array sampling ASIC is used for waveform recording at a rate up to \SI{3.0}{GSa/s} with an adjustable record length up to $\sim$\SI{700}{ns} and the capability for multi-event buffering on-chip \cite{lab4d}. For RNO-G it is planned to operate the LAB4D in 2x 2048-sample buffers for essentially deadtime-less performance.

A trigger decision can be made using input from the primary neutrino trigger board (phased-array) or an auxiliary on-board trigger using similar Schottky diode detector circuits. The auxiliary on-board trigger is formed using a comparison between a DC voltage level and the enveloped waveform, which is fed to the on-board FPGA to build a combinatoric trigger decision. As the auxiliary trigger will have a higher overall threshold than is possible with the primary neutrino trigger board, it will predominately be used as additional trigger for the surface antennas as an air shower trigger. In periods in which the power available to the stations is low (see Sec.~\ref{sec:power}) it can serve as main trigger, however, with a much weaker sensitivity to neutrino signals. 

Once an event is digitized, the waveforms and metadata are transferred to a BeagleBoneBlack Industrial, an ARMv7l Linux system, over a Serial Peripheral Interface (SPI) link, which allows data transfer at up to \SI{20}{Mbps}. 
The operating system and acquisition software are stored on robust eMMC storage, while a \SI{128}{GB} industrial SD card stage data before it is transmitted wirelessly to Summit Station. The acquisition software is an evolution of field-proven ARA phased array acquisition software.

\subsection{Autonomous power and wireless communications}
\label{sec:power}
\begin{figure}
    \centering
    \includegraphics[width=0.6\textwidth]{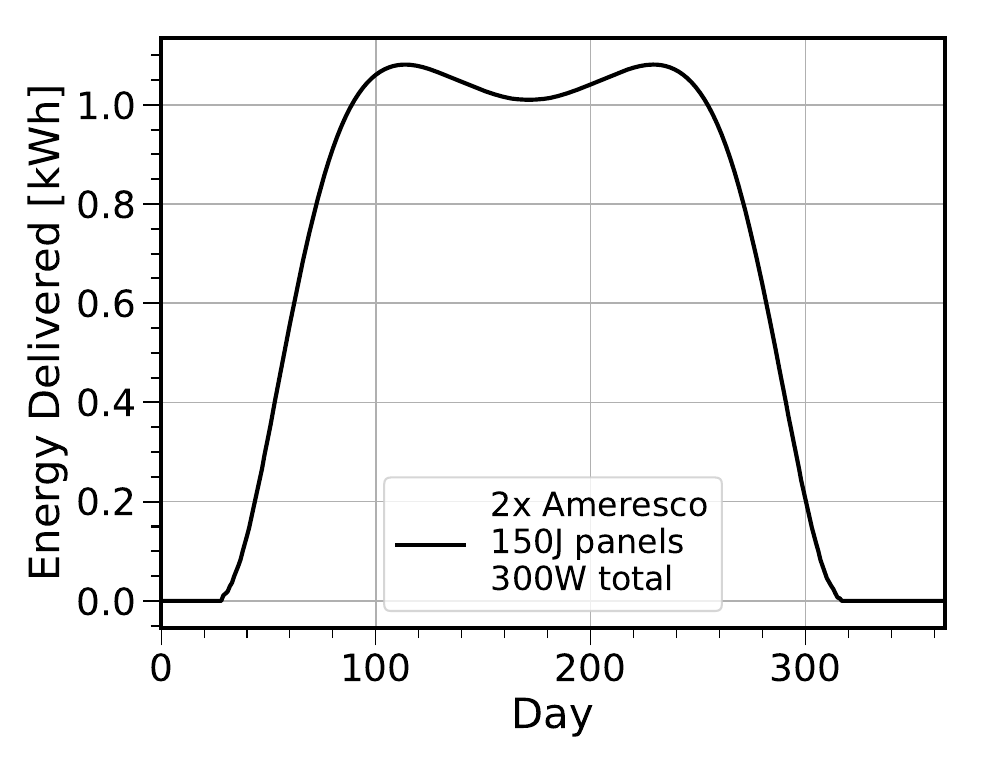}
    \caption{Predicted daily energy delivered by a \SI{300}{W} photo-voltaic (PV) array to an RNO-G station at Summit Station. The PV array comprises two Ameresco 150J rugged panels mounted vertically and facing south. The total PV area is \SI{2}{\metre\squared}. }
    \label{fig:modeled_solar}
\end{figure}

Autonomous power and wireless communications simplify logistics for an experiment of this scale and become even more efficient for even larger arrays, such as IceCube-Gen2. Each station will be powered by two solar panels, with a total maximum power output of \SI{300}{W}, and a \SI{5}{kWh} sealed lead-acid battery bank that provides three days of full-system (\SI{24}{W}) running capacity during cloudy or inclement conditions, with a 60\% de-rating margin. Lead-acid batteries, when lightly discharged relative to total capacity, have a proven track record in Arctic environments as demonstrated by the UNAVCO remote stations \cite{unavco}. 
The daily solar energy delivered to a RNO-G station using a \SI{300}{W} solar panel array is shown in Fig.~\ref{fig:modeled_solar}, using realistic estimates of 70\% total sun fraction (including diffuse and snow-reflected contributions) and a 90\% charge-controller efficiency. A low-power microcontroller ($\mu$C) will manage the power system and turn parts of the detector on and off as necessary. The $\mu$C communicates with the Beaglebone SBC via a serial connection so that the SBC may be shut down cleanly if necessary. Enough power granularity is available to run the detector in a low-power, lower-sensitivity mode if needed.

The RNO-G station can be operated in several different modes depending on the available solar power capacity, in order to maintain constant science data during long stretches of inclement weather and during the shoulder seasons, when the sun only rises above the horizon for short periods per day.  
These operating modes include: 
\begin{enumerate}
    \item {\bf Full-station mode}:  Power, trigger, and data acquisition on the full 24-channel station including the low-threshold trigger and full LTE data telemetry. Power:$\sim$24~W. 
    \item {\bf High-threshold mode}:  Power, trigger, and data acquisition on the full 24-channel station without the low-threshold trigger and minimal LTE data telemetry. Power:$\sim$17~W. 
    \item {\bf Surface-only mode}:  Power, trigger, and data acquisition only on the 9~surface LPDAs and minimal LTE data telemetry. Power:$\sim$6~W. 
    \item {\bf Winter-over mode}:  Operating mode during the polar night. All power is turned off except to the charge-controller, LoRaWAN network, and station-control microcontroller. Only minimal housekeeping data is telemetered over LoRa. The estimated power draw is $\sim$70~mW. 
\end{enumerate}

The expected uptime for an RNO-G station at Summit Camp with the \SI{300}{W} PV panel array is 216 days in operating mode 1 (59\%), 25 days in mode 2 (7\%), and another 20 days in mode~3 (5\%) for a total science livetime of $\sim$70\% averaged over the year. For the remaining 30\% of the year, the station will be put in winter-over mode. These different operating modes can be engaged by the RNO-G station controller autonomously or commanded remotely over one of the wireless networks.

\begin{figure}
    \centering
    \includegraphics[width=0.7\textwidth]{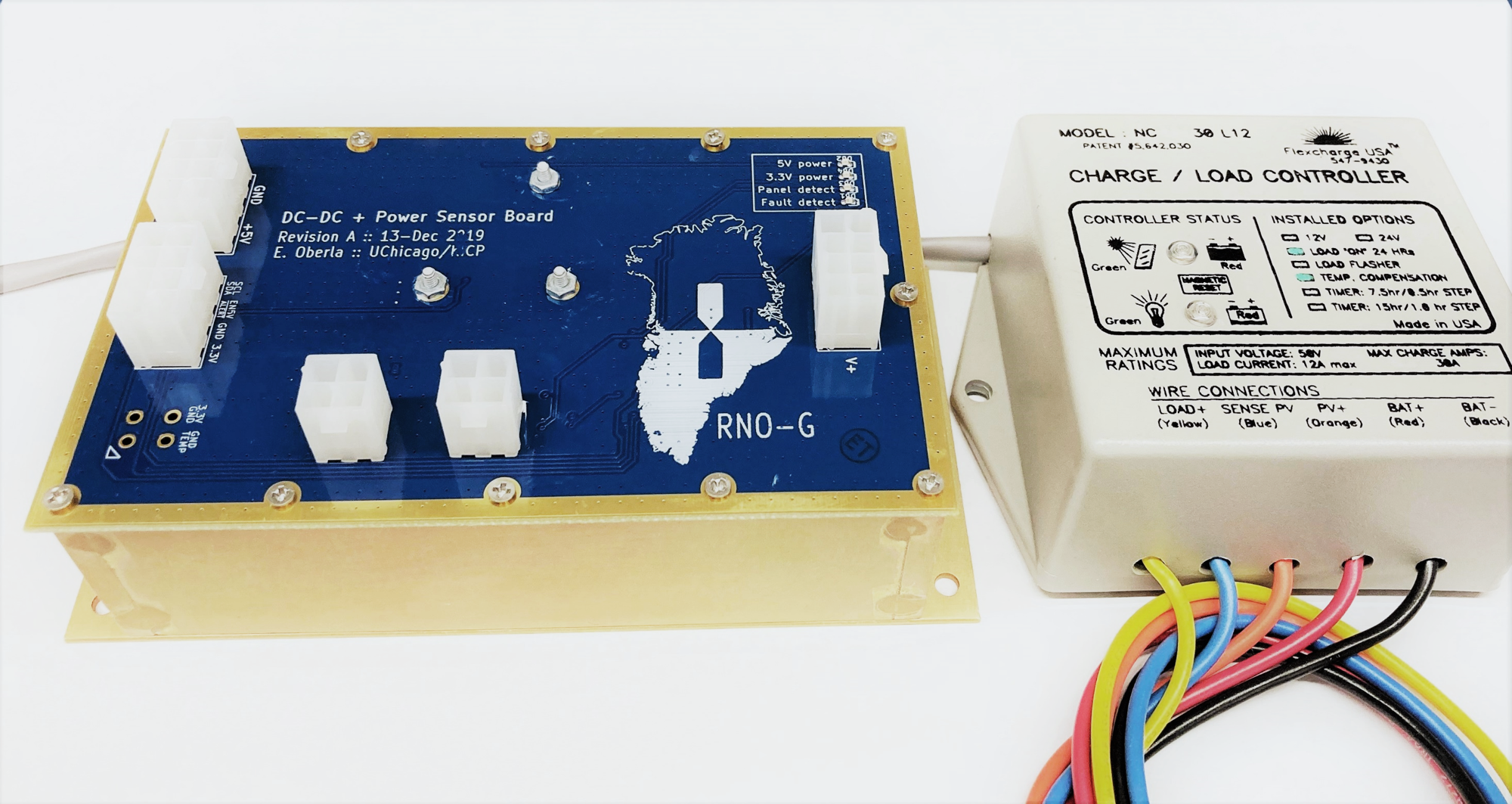}
    \caption{The station solar charge controller and high-efficiency DC-DC board for RNO-G.  
    }
    \label{fig:power_comms}
\end{figure}

Options to operate further into the winter are being explored. This R\&D is particularly relevant for a potential larger array at the South Pole such as IceCube-Gen2, where the polar night is longer. Although not part of the baseline RNO-G design, wind-turbines may allow to extend the full-station mode operations of RNO-G throughout the winter. Development of radio-quiet wind turbines that can survive in the polar environment is ongoing \cite{WindICRC}. Modeling using historical wind data \cite{historical_wind_data,noaa_weather_data} suggests that a feasible 25\%-efficient turbine at a height of \SI{10}{m} would produce a daily average of \SI{1200}{Wh} per square meter of collection area. Due to extended periods of low wind speeds a larger battery buffer will be needed for operation on wind power. 

The main data transfer link from each detector to Summit Station will use modern cellular technology. A private LTE network provides high bandwidth (up to \SI{75}{Mbps} total uplink) and long range while consuming minimal power ($<$1~W average) at each station.  A commercially-sourced LTE base station will be deployed with an antenna on the roof of the Science and Operations Building at Summit Station.  As a compromise between range and minimizing interference with our detectors, LTE Band 8 (880-915 MHz uplink, 925-960 MHz downlink) was chosen and a permit has been acquired from the Greenlandic Radio Administration.  Link modeling, including terrain shielding and a \SI{10}{dB} fading margin, predicts a usable range up to \SI{10}{km}. 

A 34-dBi roof-top sectorial antenna at Summit can cover the azimuthal extent of the array and each station will be equipped with a \SI{9}{dBi} antenna on a \SI{3}{m} mast. A secondary LoRaWAN \cite{lorawan} network will also be deployed, providing a backup low-power but low-bandwidth connection for control and monitoring.

\section{Installation, calibration, and operations}

\begin{figure}
\centering
\includegraphics[width=0.9\textwidth]{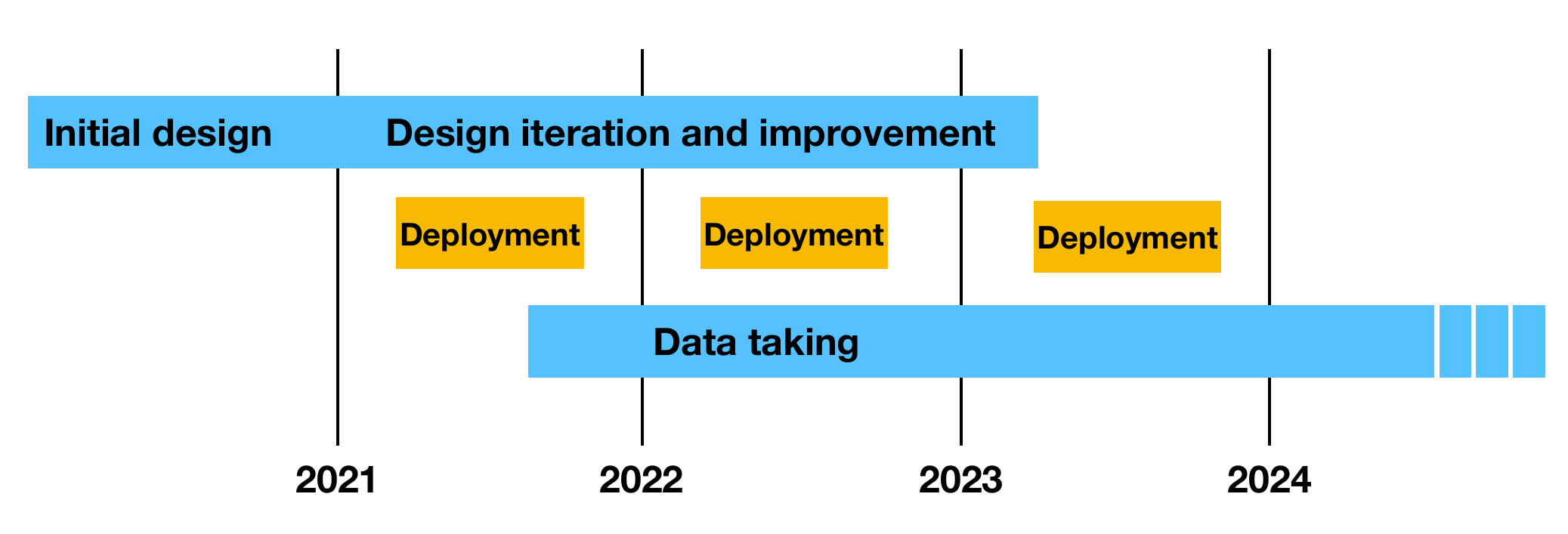}
\caption{The anticipated timeline of RNO-G. The initial design work is ongoing. Installation will take place in the summer of 2021, 2022, and 2023, tentatively scheduling the installation of 10, 10 and 15 stations, respectively. Data taking will commence with the first deployed station.   
} 
\label{fig:schedule} 
\end{figure}

The anticipated timeline of the construction of RNO-G is shown in Fig.~\ref{fig:schedule}. The initial design work is already on-going and a first installation of stations is anticipated for 2021, provided that there are no continued restrictions due to the COVID-19 virus. 

\subsection{Drilling and installation plan}
\label{sec:deployment}
The main tasks for installation of each RNO-G station are: 
\begin{enumerate}
    \item drill boreholes for deep instrumentation,
    \item deploy the solar panels and communications,
    \item deploy detector instrumentation in boreholes and trenches,
    \item confirm station operation and take calibration data.
\end{enumerate}
The baseline RNO-G scenario assumed use of the ASIG mechanical drilling technology. The ASIG drill, owned and operated by the US Ice Drilling Program (IDP) is an auger with add-in drill sections. One \SI{100}{m} deep hole requires a single working shift of 10 hours for three people.  Therefore, the three holes required for each RNO-G station can be drilled in three days assuming one work shift per day, or one and a half days assuming two work shifts per day. 

The preferred drill under consideration is the Rapid Access Isotope Drill (RAID) from the British Antarctic Survey (BAS). Holes of the diameter of \SI{3}{''} were successfully drilled to \SI{461}{m} at Little Dome C. For RNO-G larger diameter holes are needed, which is why an existing proto-type development BigRAID is being considered \cite{BAS}. It will provide  \SI{285}{mm} or \SI{11.2}{''} holes, taking about 0.85 days to reach \SI{200}{m} or 0.38 days to reach \SI{100}{m}, making it both faster and more versatile than the ASIG drill. 

Using a mechanical drilling approach is much more scalable than previous drilling efforts for the ARA experiment at the South Pole, which used a hot water drill to reach \SI{200}{m} depths. Mechanical drills are significantly lighter weight and less complex. Future development in drilling technology may enable exploring a wider range of more aggressive designs with RNO-G, which may lead to further improved sensitivity or event reconstruction capability.  Drilling below the firn layer may provide significant increases in field-of-view due to fewer limitations in ray bending.  However, care needs to be taken that any drill remains fast enough so as not to be the rate limiting step in installation and that personnel to operate the drill remains limited. Partly autonomous drilling operation is also under consideration. 

Although subject to considerations such as firn thickness (which impacts drill depth) and ice temperature, local snow accumulation rate, average daily temperatures and the availability of solar and wind power, the station design is purposely general. This allows easy adaptation of the design for future larger in-ice arrays at other sites, such as IceCube-Gen2 at the South Pole.

\begin{figure}
    \centering
    \includegraphics[width=\textwidth]{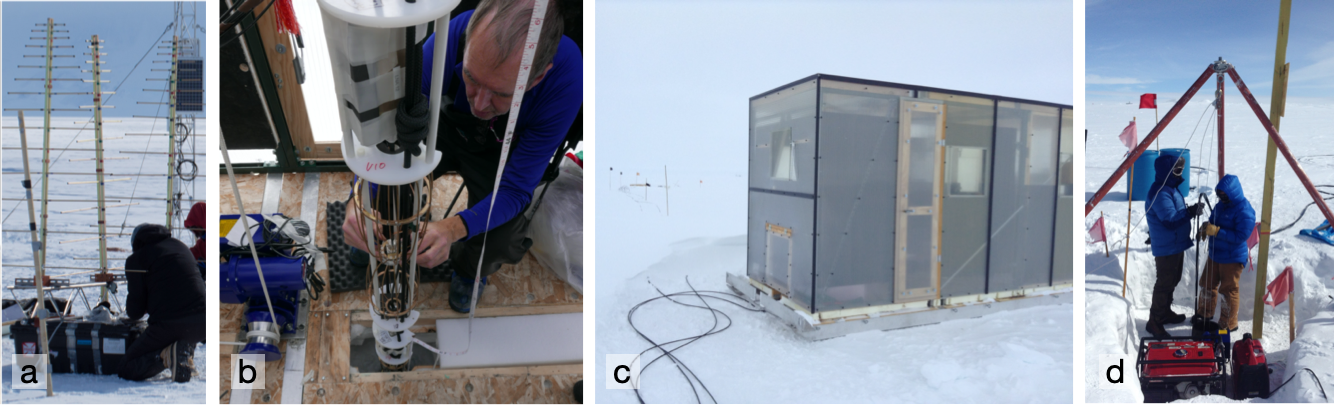}
    \caption{RNO-G installation plans are based on prior deployments of (a) surface stations at ARIANNA and (b) strings of antennas deployed in boreholes for ARA, both in compact phased arrays and reconstruction strings. A deployment shed for the drill and installation will be built on skis based on prior work done for ARA (c). Site studies conducted at Summit Station in Greenland also informed the installation plans (d).  } 
    \label{fig:drilling}
\end{figure}

The installation of both infrastructure (solar panels and communication antennas) and instrumentation is anticipated to be faster than hole-drilling. A drilling and installation team of seven people is foreseen for the first installation season, with installation beginning a week after commencing drilling. We project that an installation of up to 20 stations a year at Summit Station seems feasible. After installation, additional time will be required in the field to commission and validate station operation.

\subsection{Calibration requirements and strategies}
\label{sec:calibration}
In order to optimally reconstruct events, the relative antenna positions must be known to a small fraction of the wavelength. 
Calibration using a local radio transmitter is necessary to achieve the required few-cm precision. Two deep transmitting antennas will be included with each station, as well as one at the surface. The calibration signal is generated in the surface controller board and sent downhole over RFoF.  By measuring the relative time delays of the signal at each receiving antenna, the positions may be determined. After initial calibration, occasional runs of the pulser serve as a check of system stability. 

An existing \SI{740}{m} deep nearby borehole (DISC) \cite{johnson_2007}, will also be used to send pulses to the array from various depths. This serves as a check of the antenna position and, by varying the depth of transmission, allows inversion of the radio properties of the ice. Understanding the refractive index profile of the firn is key to  reconstruction and sensitivity modeling.  Additional pulsing from the surface will be performed to further understand the ice. 

Every station is equipped with a GPS, which will be used to synchronize event timing between stations at the \SI{10}{ns} level. This is especially important for analyzing extensive air shower events, multi-station neutrino events and for absolute time difference measurements useful for ice studies. The GPS will also track the movement of station locations with the ice flow, which will provide valuable input for ice-modeling. A higher timing precision between stations can be obtained if suitable transmitters are identified at site \cite{Corstanje:2016mia} or through the usage of airplane signals \cite{PierreAuger:2016zxi}. This may also allow the combined reconstruction of neutrino signals detected in multiple stations \cite{Garcia-Fernandez:2020dhb}, which would then yield improved precision. 

All S-parameters of amplifiers, cables, and components will be calibrated before installation. Experience from radio air shower arrays has shown that a measurement of all individual components, including a temperature-dependent gain correction will be crucial to reduce systematic uncertainties. A continually-updated MongoDB database fully integrated with the simulation and reconstruction software \cite{NuRadioMC, NuRadioReco} will be used to track the parameters of all components.

\subsection{Operations and data systems}

The acquisition software on the Single-board computer (SBC) adjusts the trigger thresholds to maintain as fast a trigger rate as possible ($\mathcal{O}$(10\,Hz)) without incurring significant deadtime. This high sustained rate drives system performance downstream, so second-stage filtering is applied on the SBC to reduce the rate of saved triggers to a time-averaged 1 Hz. Additionally, \SI{0.1}{Hz} of forced-trigger data will be recorded at regular intervals to help characterize the noise environment. 

The on-disk compressed size of each event is an estimated \SI{30}{kB}, implying an average data rate of around \SI{260}{kbps} per station at \SI{1.1}{Hz}. The LTE network can easily accommodate this rate with a relatively low duty-cycle at each modem, thereby saving power.  This rate allows storage for six weeks on the local SD cards in the event of an unexpected network outage. If more time is needed, the station can be instructed via LoraWAN to reduce the rate. In the unlikely case of simultaneous LTE and LoraWAN failure, the software on the station will automatically throttle the rate. Once data is transmitted to Summit Station, it will be stored on a redundant disk array for collection each summer. At the estimated 1 TB/station/per year of data, full build-out requires a redundant storage capacity (with margin) of \SI{35}{TB}, which can easily be achieved with a single commodity rack server (e.g. Dell PowerEdge R7515) . 

All instrument status data and event metadata as well as a subset of the waveform data (5 GB/day total) will be transmitted with low latency via Summit Station's satellite link to the University of Wisconsin for monitoring and quality assurance. A small portion of available bandwidth will be reserved for remote login for any configuration changes or remote maintenance required.  The JADE software \cite{JADE} successfully developed and deployed for IceCube data management will be used for RNO-G. For data acquisition performance, all data is initially stored in a compressed packed-binary format resembling the in-memory format used by the data acquisition system. Converters will be maintained from the raw data format to more convenient archival formats (e.g. HDF5). 

All low-latency data will be readily available to the collaboration via an interactive monitoring web site\footnote{Based on \url{https://github.com/vPhase/monutor}}. A comprehensive set of checks on the metadata and system health will be performed by the computer systems at Summit Station. Any anomalies will result in an email alert.

Monitoring duty will be apportioned to institutes on a rotating basis. While monitoring, an institution is responsible for timely investigation of all alerts and daily checks of the low-latency data for potential issues.  Weekly monitoring reports will be issued to provide historical context for any issues that may arise. 

Several mock stations, taking pure thermal noise data from terminated amplifiers, will be operated at collaborating institutions. These provide a testing ground for any configuration changes, assist with training, and help debug any issues that may arise. The pure thermal noise data also serves as a useful tool in developing analyses.


\section{Projected sensitivity of RNO-G}

In order to calculate the sensitivity of RNO-G, we have simulated the full 35-station array with a detailed modelling of the baseline hardware. Simulations for radio detectors are constantly evolving, incorporating experience from air shower simulations \cite{Huege:2013vt,AlvarezMuniz:2011bs,Abreu:2011fb,Schellart:2013bba} and previous codes for neutrino radio detectors \cite{Allison:2014kha,ThesisKamlesh,Cremonesi:2019zzc,BenICRC2019}.

All simulation results presented herein have been performed with the NuRadioMC code \cite{NuRadioMC}. For the same emission model, ice model and detector quantities, the results of this code have been shown to agree to the percent level with previous and independent codes, both for single event signatures as well as for the calculations of effective volumes. It has been found that the trigger-level sensitivities are in particular affected by the precise implementation of the trigger, the exact frequency band of the detector, the noise temperature of the system, the chosen emission model describing the Askaryan effect, whether a complete array is simulated or the array is scaled up from one station (impacting the number of events detected by multiple stations), and whether the interactions of secondary particles (taus and muons) are included in the sensitivity calculation. The latter three factors are most significant, with variations up to 50\% in effective area depending on the energy. Since, in the design process, many of the instrument parameters are not completely fixed, we carefully quote in the following the assumptions made for the array and the hardware, bearing in mind that these design sensitivities are subject to change as the instrument design matures. 

For the simulations, we use as a simplified proxy for the trigger in Sect.~\ref{sec:triggering}, a single vertical dipole per station with an amplitude threshold. A range of thresholds was used from $1.5\sigma_\textrm{noise}$ to $2.5\sigma_\textrm{noise}$ to account for possible variations in the exact design of the phased-array. Currently,  $2.0\sigma_\textrm{noise}$ is the expected to be the best proxy for the phased-array trigger using 4 dipole antennas that is in production (see Fig.~\ref{fig:phasedTrig}). Dipoles are simulated at \SI{100}{m} of depth, roughly at the same depth as the planned phased array.

We have simulated the response of a dipole of \SI{50}{cm} length similar to the one in Fig.~\ref{fig:antennas} and used it for the sensitivity calculation. The simulations performed with XFdtd \cite{xfdtd} provide full gain and phase information as a function of incoming signal direction. 

We have used NuRadioMC \cite{NuRadioMC} with the \emph{ARZ2020} parameterization given in \cite{ARZ, ARZ2020} as our model for signal emission. We have included triggers induced by secondary particles produced by the outgoing lepton after a charged current (CC) interaction, following the procedure outlined in \cite{Garcia-Fernandez:2020dhb}. The simulated station layout is that shown in Fig.~\ref{fig:layout}, with 35~stations having \SI{1}{km} spacing between them on a rectangular grid.

We first discuss the sensitivity of RNO-G to a diffuse neutrino flux and how the neutrino energy will be determined, then its angular sensitivity and lastly the sensitivity to a transient event. We will also briefly report on the expected sensitivity to air shower signals.

\subsection{Sensitivity to diffuse flux}
\label{sec:diffuse}

 \begin{figure}
    \centering
    \includegraphics[width=3.55in]{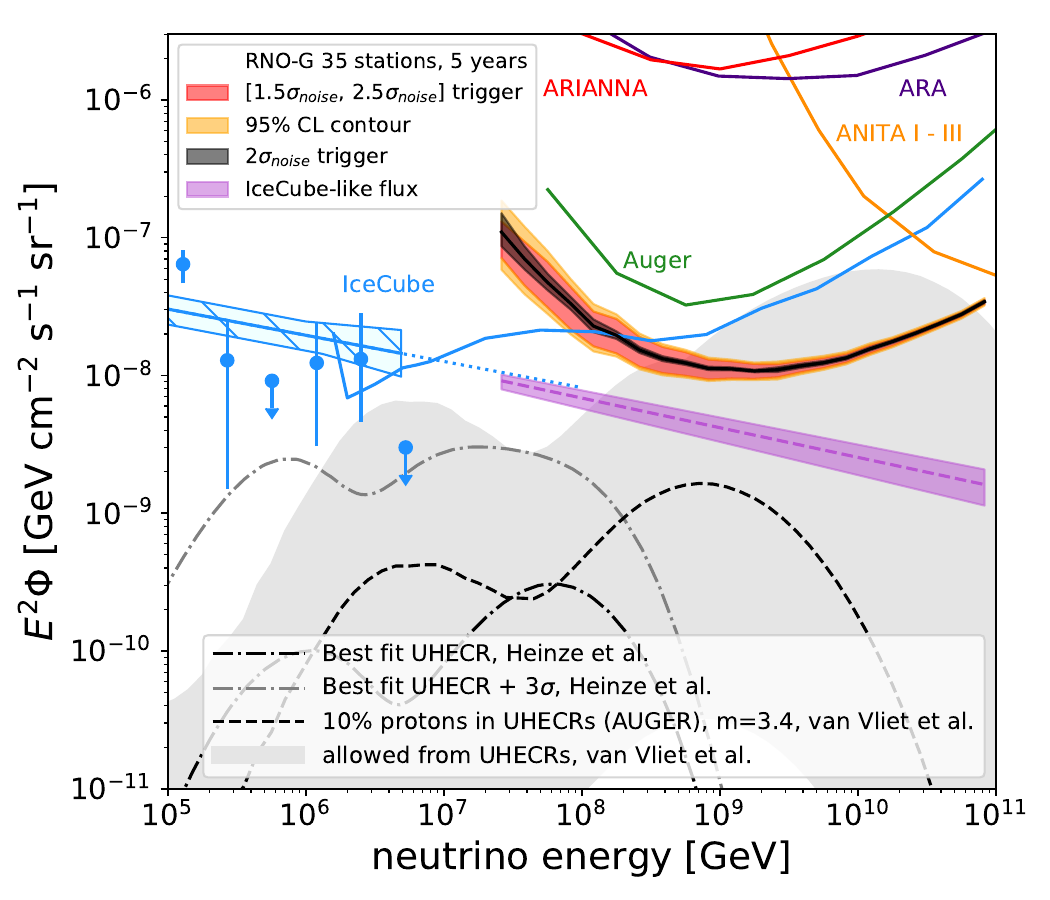} 
    \caption{The five-year sensitivity (90\%~CL upper limits) of RNO-G to the all-flavor diffuse flux for 35 stations (assuming the stations are active two thirds of the total time), compared with existing experiments and several predicted fluxes~\cite{Aab:2015kma,anita3,ara_2station,8yearnumuflux}. The red band represents the differential sensitivity band for a range of phased array proxies, spanning the interval from $1.5\sigma_\textrm{noise}$ to  $2.5\sigma_\textrm{noise}$ using decade energy bins. 95\%~CL contours are represented by the orange band. The black band is the sensitivity expected for a $2.0\sigma_\textrm{noise}$ trigger, including 95\%~CL contours. The purple band depicts the expected integrated sensitivity (90\%~CL upper limits) for an IceCube-like flux, over the [$1.5\sigma_\textrm{noise}, 2.5\sigma_\textrm{noise}$] trigger range. 
    }
    \label{fig:sensitivities_RNO}
\end{figure}

Fig.~\ref{fig:sensitivities_RNO} shows the expected 90\%~CL upper limit to an all-flavor flux for 5 years of operation of the full 35 station array, assuming a 67\% duty cycle, as expected under only solar power. This is using effective volumes for an isotropic all-sky flux and full-decade energy bins. See \cite{NuRadioMC} for more details on the $\mathrm{V_{eff}}$ calculation, and the inclusion of the  interaction length to convert from $\mathrm{A_{eff}}$ to $\mathrm{V_{eff}}$.

We have applied the Feldman-Cousins method \cite{Feldman:1997qc} for no detected events and zero background. The zero background assumption is justified as a first approximation, as according to Table~\ref{tab:muon_numbers}, we expect $\sim 0.58$~detected muons over the full energy range for five years of operation time (using SIBYLL 2.3C for signal generation and a $2\sigma_\textrm{noise}$ proxy). 


The expected upper limit is shown in Fig.~\ref{fig:sensitivities_RNO} along with other experimental bounds and model predictions. The red band shows the expected range of 90\% CL upper limits for noise levels varying from $1.5\sigma_\textrm{noise}$-equivalent trigger (lower part of the band) to $2.5\sigma_\textrm{noise}$-equivalent trigger (higher part), and includes $95\%$~CL contours due to the effective volume uncertainty. The black band shows the obtained 90\% CL sensitivity for a $2.0\sigma_\textrm{noise}$-equivalent trigger, which is the most realistic assumption for the RNO-G experiment.
We also show in Fig.~\ref{fig:sensitivities_RNO} the sensitivity for a single power law spectrum with exponents in the range indicated by the flux observed in IceCube. The purple band represents the upper limit for the IceCube flux spanned by the [$1.5\sigma_\textrm{noise}, 2.5\sigma_\textrm{noise}$] range. The dashed line in the middle of the band is the result for the $2.0\sigma_\textrm{noise}$ trigger. 
These upper limits have been calculated using the expected number of events above \SI{20}{PeV} for a range IceCube flux spectral indices and finding that value that yields the number of events equal to the Feldman-Cousins 90\%~CL upper limit under the assumption of no background.
The median upper limit exponents for the plausible trigger range cover the interval $[-2.24, -2.19]$, with $-2.21$ being the median upper limit spectral index for a $2.0\sigma_\textrm{noise}$ trigger. If no neutrino events are detected, RNO-G will be able to exclude IceCube-like fluxes above these levels.

\subsection{Energy measurement}
\label{sec:energy}
The ability of RNO-G to measure the neutrino spectrum will depend on the accuracy at which the energy of each event can be determined. 
The relation between the neutrino energy $E_\nu$ and the amplitude $|\vec{E}|$ of the electric field of the radio signal at the station is given by:

\begin{equation}
    |\vec{E}| \sim E_\nu \cdot y \cdot f(\varphi) \cdot \frac{\exp(-d/l_{\mathrm{atten}})}{d}
    \label{eq:energy_reconstruction}
\end{equation}
where $y$ is the fraction of the neutrino energy deposited into the shower, and $f(\varphi)$ a dependence on the angle under which the particle shower is observed. The last term accounts for the attenuation of the radio signal as it travels to the antenna, with $d$ being the distance of the interaction vertex from the station and $l_{\mathrm{atten}}$ the attenuation length of the ice.

In general, the (\emph{inelasticity}) fraction $y$ of the neutrino energy that contributes to a particle shower undergoes event-by-event fluctuations and cannot be reconstructed on a single-event basis. It therefore must be estimated from theory, resulting in a statistical uncertainty of, on average, a factor of $\sim 2$ \cite{Anker:2019zcx}. This restriction imposes a hard bound on the energy resolution obtainable with any neutrino detector that only observes the cascade. The goal is therefore to reconstruct the other parameters in Eq.~\ref{eq:energy_reconstruction} precisely enough for the uncertainty in $y$ to be dominant. 
It should be noted that in case of an electron neutrino interaction, the full amount of energy is transferred to two particle showers very close to each other, which argues against having the unknown fraction $y$ as bound. However, these two cascades can interfere constructively or destructively and pure-electromagetic cascades are subject to the LPM-effect \cite{Landau:1953um, Migdal:1956tc} at high energies, which changes their radio emission \cite{PhysRevD.82.074017}. This makes it reasonable to treat the inelasticity for all cases as bound in a first general consideration. 

\begin{figure}
    \centering
    \includegraphics[width=\textwidth]{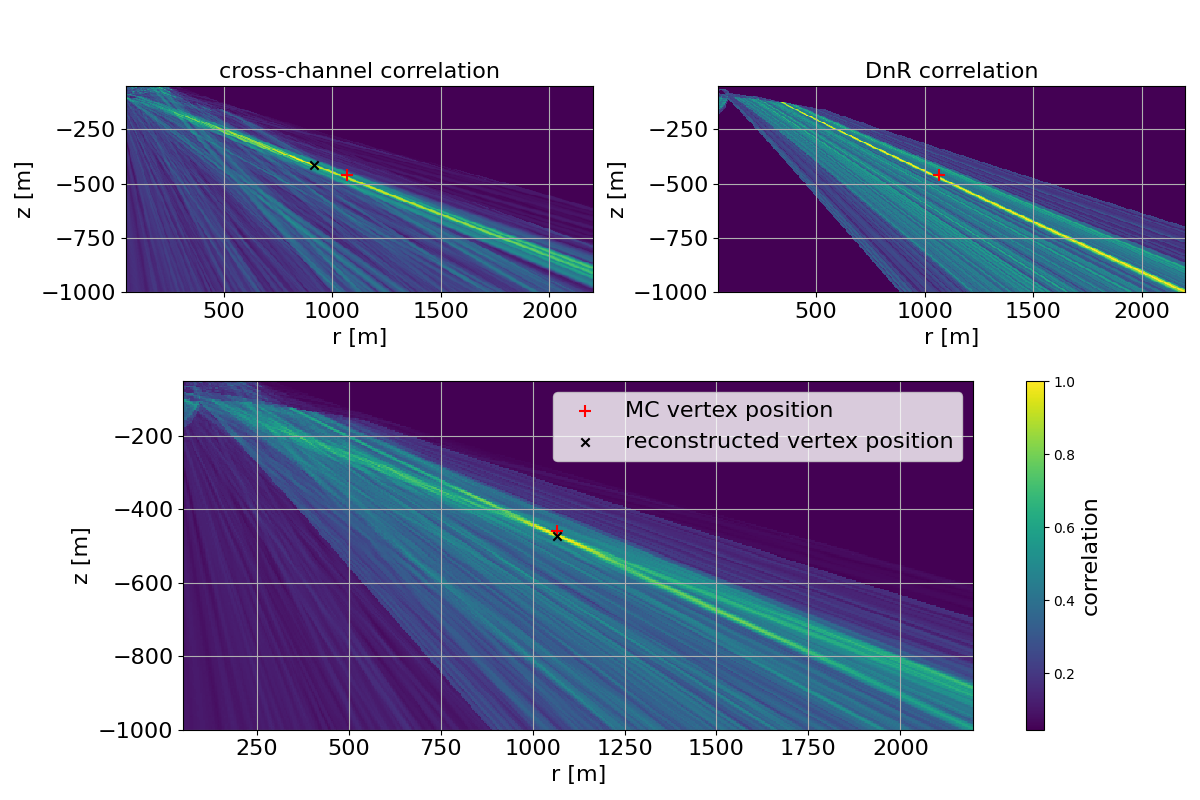}
    \caption{Neutrino interaction vertex reconstruction for one event using correlations between different channels (top left), correlations between different rays reaching the same antenna (top right) and a combination of both (bottom). Colors specify the normalized sum of correlations between channels, shifted by the difference in signal travel time expected for a given vertex position. 
    }
    \label{fig:vertex_reco}
\end{figure}

The resolution on the RNO-G measurement of the full electric field $|\vec{E}|$ depends on a number of factors. Ideally, the amplitude should be obtained for all polarization components, with separate levels of noise. In general, the larger the detected amplitude of the signals (larger measured signal-to-noise ratio (SNR$_m$), see below), the smaller the influence of noise on the uncertainty. Similarly, noise effects are mitigated as antenna hit-multiplicities increase.
As the Hpol antennas have lower gain than the Vpol antennas, the Hpol signals will typically have smaller SNR$_m$. Several methods such as \textit{forward folding} \cite{NuRadioMC}, template matching \cite{Barwick17}, or information field theory \cite{2013A&A...554A..26S} can be used to mitigate noise effects; nevertheless, the obtainable resolution of the amplitude will vary significantly from event to event. 

It should be pointed out that using SNR$_m$  differs from the situation of simulations (as defined in Sec.~\ref{sec:low_threshold}), as the true amplitude of the signal $S$ without noise is unknown, so the measured SNR$_{m}$ = (signal + noise) / noise. Using a definition of $ \mathrm{SNR}_m = 0.5 (\max(S) - \min(S))/\sigma_{\textrm{noise}}$, a typical waveform of the length of RNO-G has a roughly 50\% chance of reaching $ \mathrm{SNR}_m = 3$ simply by fluctuations of noise. At $\mathrm{SNR}_m = 3.5$ this probability is reduced to about 1\%.  

Due to constructive interference, the radio signal emitted by the particle shower is strongest if viewed directly at the Cherenkov angle, and diminishes (in a frequency-dependent manner) the further the observer viewing angle departs from the Cherenkov angle. 
As shown in Fig.\ref{fig:askaryan_pulses}, the higher frequencies lose signal coherence earliest. Therefore, the shape of the frequency spectrum of the signal can be used to reconstruct the viewing angle relative to the Cherenkov angle and, ultimately, make a correction. This method has been demonstrated for particle showers in air \cite{Welling:2019scz}, and our first simulations indicate the same to be true for neutrino showers. Quantitatively, we anticipate that $f(\varphi)$ will be obtainable for RNO-G for signals detected with at least a measured $\mathrm{SNR}_m = 3.5$.  

The signal pathlength $d$ (Eq.~\ref{eq:energy_reconstruction}) will depend on the reconstruction of the interaction vertex, so the resolution of the vertex position is another important ingredient for energy reconstruction.  

Fig.~\ref{fig:vertex_reco} shows one example of vertex reconstruction for a simulated neutrino interaction detected with RNO-G. This method to obtain the vertex position is based on cross-correlating the signals detected in all antennas with each other and deriving a probability map of the vertex location. Especially for those events in which RNO-G records both the direct emission, as well as the one reflected at-/refracted-below the surface, the resolution on the vertex position will be excellent, making the unknown factor $y$ (Eq.~\ref{eq:energy_reconstruction}) the dominating uncertainty. Further work will be carried out to determine the fraction of events for which a good vertex resolution will be obtainable and the $\mathrm{SNR}_m$ for which this will be possible. Preliminary results indicate that, conservatively, an analysis efficiency at least corresponding to the green curve in Fig.~\ref{fig:analysis_efficiencies} is reachable for the vertex and thereby energy reconstruction.  

The profile of the attenuation length of the ice in Greenland, which defines $l_{\mathrm{atten}}$ in Eq.~\ref{eq:energy_reconstruction} has been measured~\cite{avva1} and is used for the simulations. The remaining systematic uncertainty and variations across the array will be addressed by additional calibration campaigns as discussed in Sec.~\ref{sec:calibration}.

\subsection{Angular sensitivity}
\label{sec:angular_sensitivity}
\begin{figure}
\centering
\begin{minipage}{\textwidth}
  \centering
  $\vcenter{\hbox{\includegraphics[width=0.45\textwidth]{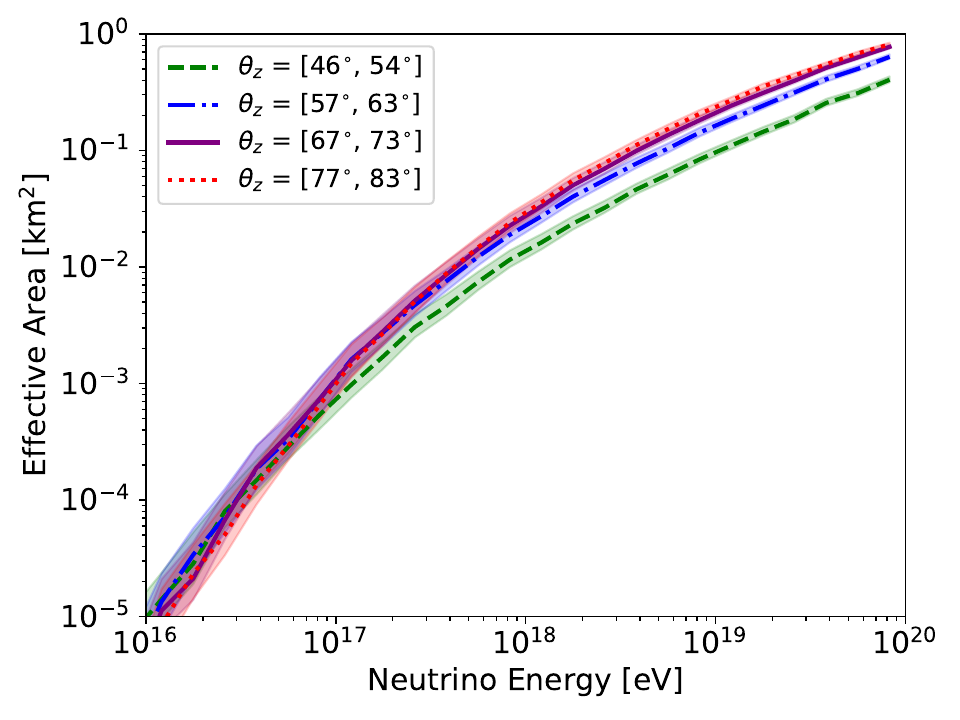}}}$
  $\vcenter{\hbox{\includegraphics[width=0.54\textwidth]{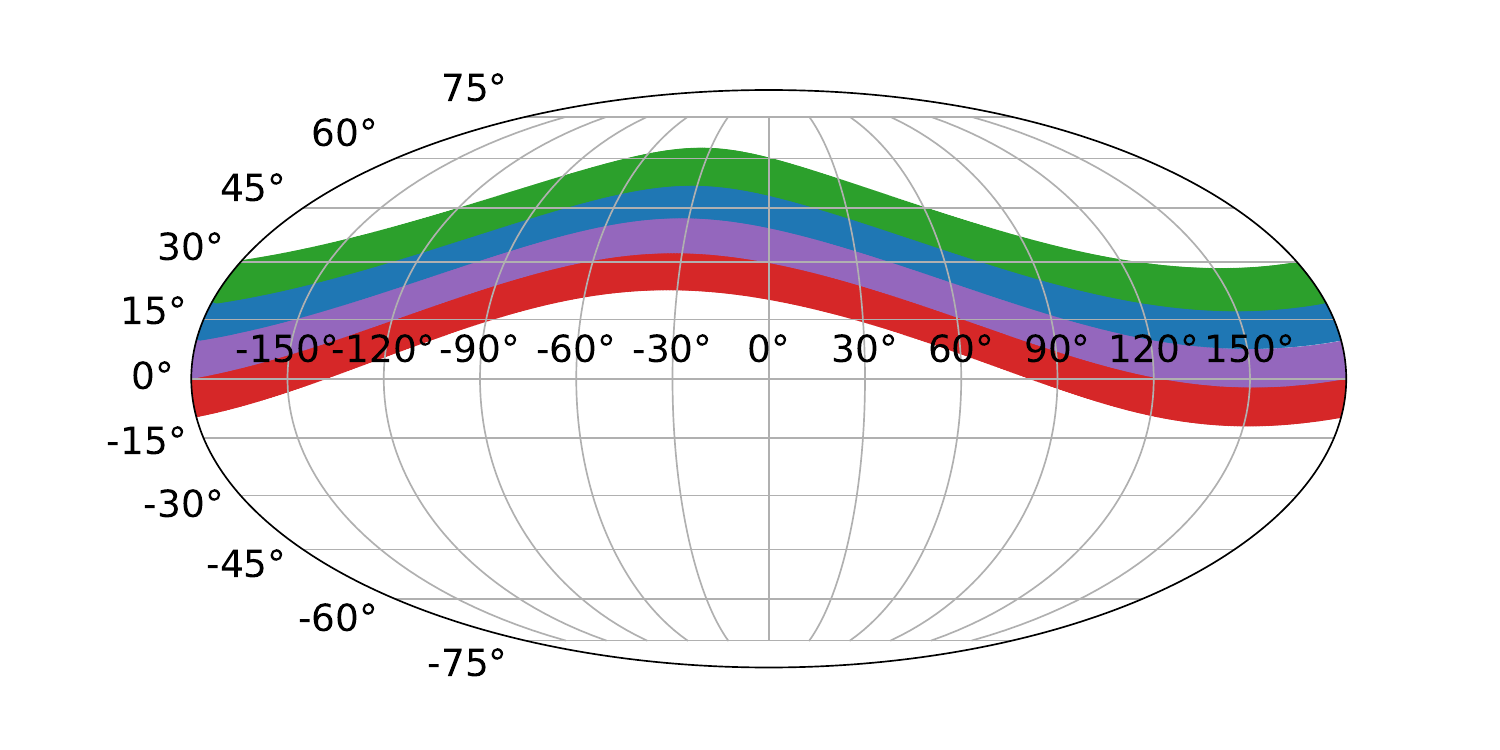}}}$
\end{minipage}
\caption{RNO-G instantaneous sky coverage. Left: Simulated effective area as a function of neutrino energy is shown for four zenith bands, centered at $50^{\circ}$, $60^{\circ}$, $70^{\circ}$, and $80^{\circ}$. Shaded regions indicate the range given by different triggers of $1.5 \sigma_{\mathrm{noise}}$ and $2.5 \sigma_{\mathrm{noise}}$.  Simulations were performed for the full RNO-G array of 35 stations with a distance of \SI{1}{km}. Right: These bands are projected in Right Ascension (RA) and Declination (Dec) for one particular time of day to illustrate the instantaneous sky coverage. For zenith angles $<45^{\circ}$ or $>90^{\circ}$ RNO-G sensitivity is strongly reduced (< 0.1 fraction of maximum effective area
} 
\label{fig:sky_coverage}
\end{figure}

The sky coverage of RNO-G is mostly determined by the geometry of its location in Greenland. In Fig.~\ref{fig:sky_coverage} we show the effective areas for different zenith angle bands for RNO-G, as well as their projection onto equatorial coordinates. Outside of these bands, the effective area decreases rapidly (see also \cite{Garcia-Fernandez:2020dhb}), making RNO-G mostly sensitive to an annulus of roughly $45^{\circ}$ just above the horizon. 

The ability of RNO-G to provide an accurate arrival direction for detected neutrinos depends on its ability to detect the signal arrival direction and the angle with respect to the Cherenkov cone, as well as the signal polarization, and is again a strong function of the number of antennas with detected signal and their $\mathrm{SNR}_m$. 

\begin{figure}
    \centering
    \includegraphics[width=0.8\textwidth]{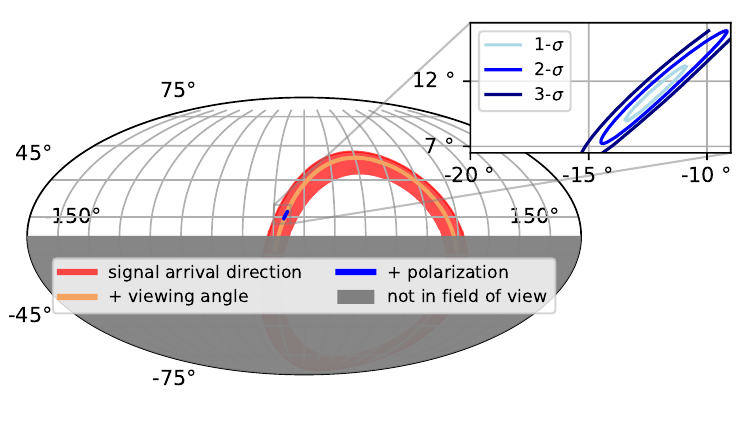}
    \caption{Radio neutrino detector arrival direction reconstruction. Given the limited field of view, the reconstructed signal arrival direction restricts the neutrino arrival direction to the red circular band shown. Adding information from the frequency content constrains the viewing angle, and reduces the width of the band. Finally, including data from both Hpol and Vpol antennas, polarization reconstruction reduces the allowed arrival direction to a small area on sky. The reconstruction and uncertainties are shown for one event simulated for RNO-G with NuRadioMC. The coordinates are local azimuth and zenith angle. 
    }
    \label{fig:angular_resolution}
\end{figure}

The signal arrival direction can be directly determined from the time difference in the captured channel-by-channel waveforms, using (for example) cross-correlation. The obtained resolution is a function of the number of antennas with signal; sub-degree values have typically been obtained by previous experiments \cite{ara_2station,glaciology-draft,Anker:2019zcx,Anker:2020bjs}.
Knowing only the arrival direction for the signal at a specific station, the neutrino arrival direction can be determined to lie on a cone, projecting to a ring on-sky as shown in Fig.~\ref{fig:angular_resolution}. Only a fraction of the ring corresponds to a probable physical solution, as many arrival directions can be excluded by the known Earth absorption. 

The radio signal is the strongest on the Cherenkov cone and then weakens once the angle to the shower axis deviates from the Cherenkov angle. Depending on the type of event, viewing angles of more than 10 degrees with respect to the Cherenkov angle may still be observable.
As discussed in Sec.~\ref{sec:energy}, the electric-field is a function of the viewing angle, as the higher frequencies fall off further away from the Cherenkov cone, so the viewing angle is reconstructable via the frequency slope. Combining signal arrival direction and viewing angle narrows the ring of possible arrival directions. 

As the radio signal is due to the Askaryan effect, the polarization of the induced electric-field points radially inwards towards the shower axis. Therefore, a measure of the polarization is needed for a unique neutrino arrival direction. As shown in Fig.~\ref{fig:angular_resolution}, adding polarization allows reducing the entire ring to a small patch on the sky. The absolute angular resolution as function of energy, elevation and $\mathrm{SNR}_m$ per antenna is still under study. Thus, Fig.~\ref{fig:angular_resolution} has been constructed to highlight the influence of different signal parameters on the angular resolution, while using a simulated event, as detectable in RNO-G, including noise but no detector uncertainties. The event shown has an $\mathrm{SNR}_m\approx 6$ in both Vpol and Hpol antennas, meaning that all pulses can be clearly identified.

\subsection{Sensitivity to transient events}
\begin{figure}
    \centering
    \includegraphics[width=4in]{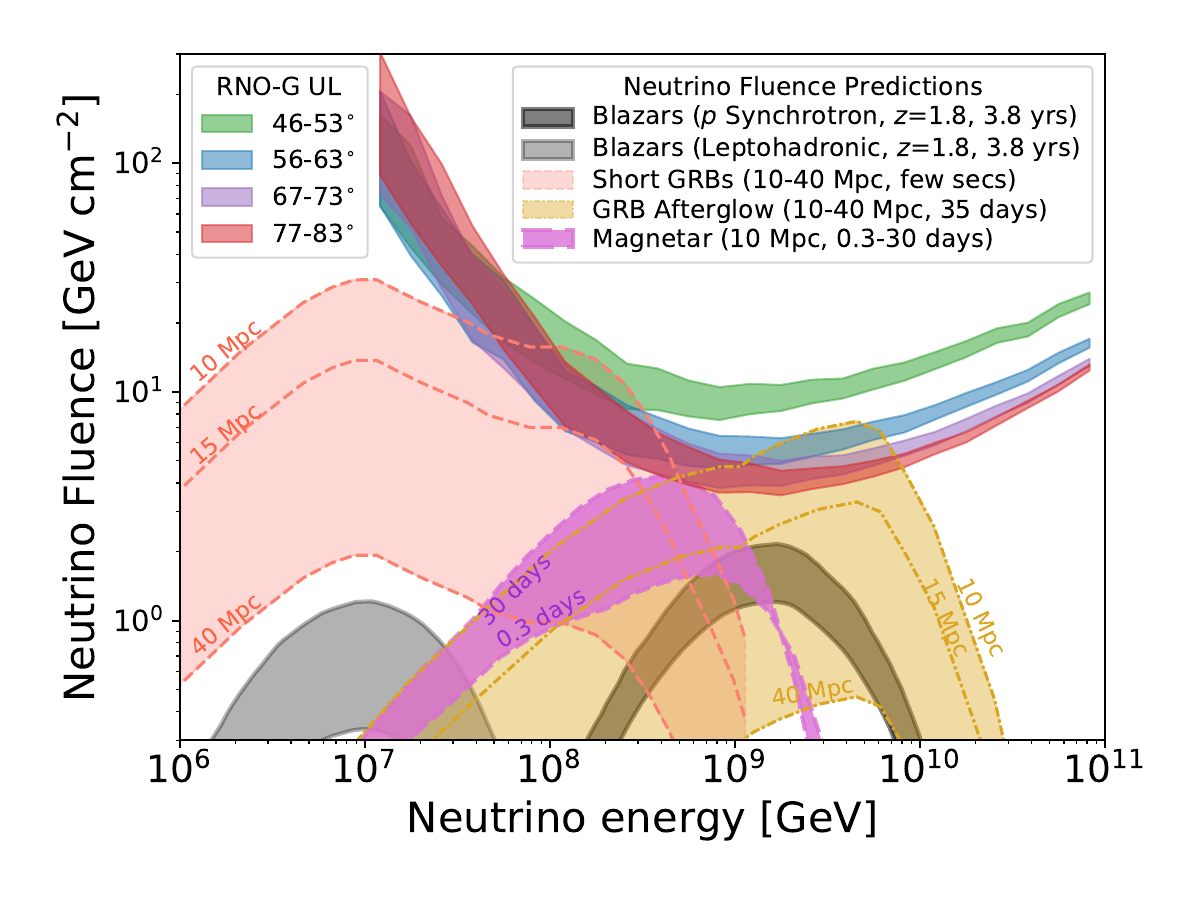}
    \caption{95\% CL fluence sensitivities between triggers at $1.5 \sigma_{\textrm{noise}}$ and $2.5 \sigma_{\textrm{noise}}$ are shown for four zenith bands centered at (top to bottom) 50$^\circ$ (green), 60$^\circ$ (blue), 70$^\circ$ (purple), and 80$^\circ$ (red). Sensitivities are calculated for a full decade in energy. Model-predicted fluences from several transient classes (bright gamma-ray blazars \cite{rodrigues2020multi}, short GRBs \cite{Kimura:2017kan}, magnetars~\cite{Fang:2017tla}, and GRB afterglows \cite{Murase:2007yt}) are also shown for direct comparison. We scale the short GRB and GRB afterglows by several luminosity distances to demonstrate the distance over which RNO-G will be sensitive to transients; a similar scaling can be applied to other source classes. For the calculation of sensitivities here we have used an integrated background expectation of no events. Note that for longer duration transients, integrated background may become non-negligible.  
    }
    \label{fig:transients}
\end{figure}

Using the same simulations as performed for Sec.~\ref{sec:diffuse}, the sensitivity of RNO-G to transient events has been obtained, as shown in Fig.~\ref{fig:transients}. Most models predict small neutrinos fluxes in the energy range of RNO-G, as compiled in Fig.~\ref{fig:duration}. However, given, e.g., large uncertainties in the modelling of mergers of neutron stars and that this area of multi-messenger astronomy is still in its infancy, RNO-G may make serendipitous discoveries. Its location in the Northern hemisphere makes it uniquely sensitive, and complementary to other planned radio neutrino observatories in the Southern Hemisphere. 

GRBs and other cataclysmic events are promising candidates for transient flares of UHE neutrinos. 
GRB afterglows are expected to produce the highest energy neutrinos over months-long time scales~\cite{Murase:2007yt}.  Short GRBs resulting from binary neutron star mergers may be detectable with RNO-G if they are nearby or connected with the production of giant flares from magnetars \cite{Yang:2020xwc}. Similarly, magnetars resulting from binary neutron star mergers can drive UHE neutrino production~\cite{Fang:2017tla}. As shown in Fig.~\ref{fig:transients}, RNO-G can constrain the neutrino fluence from GRB afterglows, short GRBs, and long-lived magnetars within tens of Megaparsecs. Furthermore, Tidal Disruption Events (TDEs) are another cataclysmic source class still in the infancy of their discovery, with frequent new observations and population increases thanks to transient observatories such as the Zwicky Transient Facility (ZTF) \cite{van2020seventeen}. As more is uncovered about their nature, they may also become a viable multi-messenger target for RNO-G. 

Flaring blazars are particularly interesting targets for RNO-G. 
As an example, a model of the neutrino fluence expected from the flare of the bright gamma-ray blazar PKS 1502+106 \cite{rodrigues2020multi} is compared to the RNO-G sensitivities in Fig.~\ref{fig:transients}. This particular blazar is an FSRQ, which are notable for their expected high UHE neutrino fluxes \cite{righi2020eev}, and spatially coincident with a ``golden'' event (IC190730A) seen in IceCube \cite{2019GCN.25241....1I,Franckowiak:2020qrq}. In the model, neutrinos are produced in the two different scenarios that are consistent with multi-wavelength photon observations, but the neutrino spectrum is strongly impacted by the radiation mechanism. Stacking searches in RNO-G for flares of blazars or multi-messenger driven searches may reveal UHE neutrinos or constrain the neutrino spectrum at the highest energies. Note that while PKS 1506+106 is at a distant redshift, closer blazars will have a stronger neutrino fluence.

RNO-G has unique capabilities to process alerts in nearly real time. Summit Station's continuous satellite link and the LTE communications strategy can permit alerts from other multi-messenger observatories to be sent to and from the RNO-G stations. 

\subsection{Sensitivity to air shower signals}
RNO-G will be equipped with upward-facing LPDAs, sensitive to air shower signals. These will be triggered through the auxiliary trigger as described in Sec.~\ref{sec:triggering}. First simulations indicate a turn-on of the trigger efficiency to air showers between \SI{1e16}{eV} and \SI{1e17}{eV}, with details depending on the exact system noise temperature and environmental noise conditions that will need to be confirmed during the first deployment season in-situ. The DAQ is designed to store \SI{0.1}{Hz} of triggers from the surface antennas, dedicated to the detection of air showers. The passband of the envelope trigger has been optimized for the highest surface antenna trigger efficiency and will be between \SI{80}{MHz} and \SI{180}{MHz}. We expect the detection in the order of one air shower per day per station. 

The air shower trigger at RNO-G will serve two purposes. As discussed in Sect.~\ref{sec:low_bg}, the muonic component of air showers may constitute a background for neutrino detection with RNO-G. While the flux of these background events depends strongly on the composition of the cosmic ray flux, as well as hadronic interaction models, the safest way to contain the impact of this background is to unambiguously tag air showers. RNO-G will therefore continue to be optimized to provide its own air shower veto. In addition, air shower reconstruction will help calibrate the system and ensure an independent cross-check of up-time and efficiency.


\section{Conclusions}
We have presented the concept of the Radio Neutrino Observatory in Greenland (RNO-G), currently scheduled to commence installation at Summit Station in 2021. The location in Greenland both drives design considerations, such as autonomous low-power stations, and, given the unique field of view from the Northern Hemisphere, also defines the strong science case.

The RNO-G hardware builds on previous radio array experience and strives for a very low-noise system that can sustain a low trigger-threshold, but high duty-cycle operation of autonomous stations. Each of the 35 RNO-G stations will consist of log-periodic dipole antennas deployed at the surface and custom-made dipole and tri- or quad-slot antennas deployed in three mechanically drilled holes to a depth of \SI{100}{m}. The stations will mainly be triggered by a phased array of four deep dipoles at the \SI{100}{m} maximum depth, which will ensure the best neutrino aperture. Auxiliary envelope triggers are available for low-power operations in the seasons with less abundant solar-power and for reading out the surface antennas to detect and veto air showers. 

RNO-G will be the first uniform deployment of a neutrino radio array that will demonstrate the feasibility of scaling to arbitrarily large arrays.
The delivered per-year sensitivity will be the largest achieved to-date with a radio array. RNO-G with its unique view of the Northern hemisphere may provide insights into transient sources of UHE neutrinos and will bring the detection of a continuation of the astrophysical neutrinos flux to high energies as detected by IceCube within reach. Additionally, models for cosmogenic neutrinos assuming a significant proton fraction in UHE cosmic-rays will be either be conclusively ruled out or will lead, if confirmed, to a detection of neutrinos with RNO-G. 

\section{Acknowledgements}

We would like to acknowledge our home institutions and funding agencies for supporting the RNO-G work; in particular the Belgian Funds for Scientific Research (FRS-FNRS and FWO) and the FWO programme for  International Research Infrastructure (IRI), the
German research foundation (DFG, Grant NE 2031/2-1), and the Helmholtz Association (Initiative and Networking Fund, W2/W3 Program). 

\bibliographystyle{JHEP}
\bibliography{BIB}

\end{document}